\documentclass[aps,prb,showpacs,raggedbottom,nobalancelastpage,amssymb]{revtex4}
\usepackage{amsmath}
\usepackage{amsfonts}
\usepackage{graphicx}
\usepackage{appendix}
\usepackage{dsfont}
\usepackage{amssymb}
\usepackage{bm}
\usepackage{color}
\usepackage{ulem}
\usepackage{soul}
\usepackage{natbib}
\setstcolor{red}
\usepackage{nccmath}
\usepackage{array}

\newcommand{\beq}{\begin{equation}}
\newcommand{\eneq}{\end{equation}}
\newcommand{\be}{\begin{equation}}
\newcommand{\ee}{\end{equation}}
\newcommand{\bea}{\begin{eqnarray}}
\newcommand{\eea}{\end{eqnarray}}

\usepackage{multirow}
\usepackage{wasysym}


\begin{document}
\title{Dual fermionic variables
and  renormalization group approach to junctions of strongly interacting quantum wires}
\author{Domenico Giuliano$^{1,2}$  and Andrea Nava$^{1}$ }
 \affiliation{$^{1}$
Dipartimento di Fisica, Universit\`a della Calabria Arcavacata di Rende I-87036, Cosenza, Italy
and
I.N.F.N., Gruppo collegato di Cosenza, Arcavacata di Rende I-87036, 
Cosenza, Italy  \\
$^2$ CNR-SPIN, Monte S.Angelo via Cinthia, I-80126 Napoli, Italy}
\date{\today}

\begin{abstract}
Making a combined use of  bosonization and fermionization techniques, we
 build  nonlocal transformations between dual fermion operators, describing 
junctions of strongly interacting  spinful one-dimensional quantum wires. 
Our approach allows for trading  strongly interacting (in the original coordinates) 
fermionic Hamiltonians for weakly interacting (in the dual coordinates) ones. 
It enables us to generalize to the strongly interacting regime the  fermionic renormalization group 
approach to weakly interacting junctions.  
As a result, on one hand, we are able to pertinently complement the information
about the phase diagram of the junction obtained within bosonization approach; 
on the other hand, we map out the full  crossover of the conductance tensors between any two 
fixed points in the phase diagram connected by a renormalization group trajectory. 
\end{abstract}


\maketitle
 
\section{Introduction} 
 \label{intro}

Many-electron systems, such as conduction electrons in a metal, 
are typically well-described by Landau's Fermi liquid theory (LFLT), which is  
capable of successfully accounting  for the main effects of electronic interaction 
\cite{landau,landau2,landau3,agd}. LFLT's basic assumption is that, close to 
the Fermi surface, low-energy  elementary 
''quasiparticle'' excitations of the interacting electron liquid  are in one-to-one correspondence with 
particle- and hole-excitations in the noninteracting 
Fermi gas. This corresponds to a nonzero overlap between 
the quasiparticle wave function in the Fermi liquid and the electron/hole
wave function in the noninteracting Fermi gas, which is reflected
by a coefficient $Z_F$ of the quasiparticle peak at the Fermi 
surface in the spectral density of states that is finite though, 
in general, $<1$ ($Z_F = 1$ corresponds to the noninteracting 
limit). The ''adiabatic deformability'' of quasiparticles to 
electrons and/or holes by smoothly switching off the interaction 
allows, for instance, to address transport in a Fermi liquid in 
a similar way to what is done using scattering approach in 
a noninteracting Fermi gas, etc. \cite{glazman_fisher}. 

LFLT is grounded on the possibility of obtaining reliable 
results by perturbatively treating
the electronic interaction \cite{agd,nozier_1}. This is strictly related to the 
small rate of multi-particle inelastic processes, in which, due to 
the interaction, an electron/hole emits  electron-hole pairs.
While this is typically the case in systems with spatial dimension
$d$ higher than one, in one-dimensional systems, the proliferation of particle-hole pair
emission at low energies  leads to a diverging corresponding rate, which makes the quasiparticle 
peak disappear ($Z_F = 0$). As a result, the interaction cannot be dealt with 
perturbatively and one has rather to resort to nonperturbative techniques, 
allowing for summing over infinite sets of diagrams \cite{solyom,shankar,polch}.
The breakdown of LFLT in one-dimensional interacting electronic systems
(interacting ''quantum wires'' - QW's) reveals itself in a series 
of physical contexts, the more remarkable of which is   the transport 
across  an interacting QW in the presence of a constriction, or a weak link, or 
the conductance of a junction of QWs, with or without spin 
\cite{furusaki,kane_fisher1,kane_fisher2,matgla1,matgla2,nazarov_glazman,nayak_1,chen,lal_1,pham,oca1,oca2,rao,chamon}.
Moreover, due to the remarkable correspondence between 
fermionic- and spin-bosonic one-dimensional lattice models 
encoded in the Jordan-Wigner fermionic representation of 
quantum spin-1/2 spin operators \cite{jordan_wigner}, a number 
of bosonic realizations of one-dimensional models of 
interacting fermions have been proposed in bosonic systems, such 
as quantum spin chains \cite{eggert_affleck,affleck_1,crampe_trombettoni,tsvelik_1,tsvelik_2,giuliano_1}, 
  quantum Josephson junction  networks 
\cite{glazman_larkin,giuliano_sodano5,cirillo}, or topological Kondo-type systems
\cite{beri_cooper,galpin,beri_cooper2,eriksson1,eriksson2}.

Formally, interacting electrons in one dimension are commonly treated 
within Tomonaga-Luttinger liquid (TLL)-approach \cite{tomonaga,luttinger}. 
TLL formalism provides a general description of low-energy physics of 
a one-dimensional interacting electronic system in terms of collective 
bosonic excitations (charge- and/or spin-plasmons).
Electronic operators are realized as nonlinear vertex operators of the 
bosonic fields \cite{haldane_1,haldane_2}. As for what concerns transport properties, the 
most striking prediction of the TLL-approach is possibly the power-law 
dependence of the conductance on the low-energy reference scale 
(''infrared cutoff''), which is typically identified with the (Boltzmann
constant times the) temperature, 
or with the (Fermi velocity times the) 
inverse system length (''finite-size gap'') in a dc transport
measurement, or with   $e V$, with $V$ being an applied voltage, in 
a nonequilibrium experiment. While TLL-formalism poses no particular 
constraints on the strength of the ''bulk'' interaction 
within the quantum wires, it suffers of  limitations, when used to 
describe transport across impurities in an interacting quantum wire, 
or conduction properties at a junction of quantum wires (generically referred
to, in the following of this section, as quantum impurities in the quantum wire).
 Parametrizing the spatial 
coordinate in each wire with $x (\geq 0)$, within the TLL-framework, a junction of quantum wires is 
mapped onto a model of $K$-one dimensional TLLs (one per each wire), interacting 
with each other by means of a ''boundary interaction'' localized at $x=0$. 
Dealing with such a class of boundary problems requires pertinently 
setting the boundary conditions on the plasmon fields at $x=0$. While 
in some very special cases the boundary conditions can be written as 
simple linear relations between the plasmon fields, in general they cannot.
This is a consequence of the nonlinearity of the relations between bosonic and fermionic fields: 
even linear conditions among the fermionic fields are traded for highly nonlinear
conditions in the bosonic fields. As a consequence, except at the  fixed points of the boundary 
phase diagram, where  the boundary conditions are ''conformal'', that is, linear in 
the bosonic fields, the boundary interaction can only be dealt with perturbatively,  
with respect to the closest conformal fixed point \cite{das,aris}. With very few remarkable exceptions, 
where the junction model Hamiltonian maps onto some exactly solvable model \cite{exact1,exact2,exact3}, 
the only way for recovering ''global'' (i.e., not necessarily in the 
vicinity of a conformal fixed point) informations about the phase 
diagram and the corresponding scaling of the conductance, is by just making 
educated guesses from the global topology of the fixed-point manifold of 
the phase diagram. Recently, the bosonization approach in combination with 
zero-temperature numerics has successfully been employed to compute the 
junction conductance by relating it to  the  asymptotic behavior of 
certain static correlation functions. Correspondingly, the length scale
over which the asymptotic behavior emerges has been worked out, as well 
\cite{feiguin_PRL,feiguin_PRB}. Also, numerical calculations of the finite-temperature
junction conductance can be performed by using quantum Monte Carlo approach
\cite{Sedlmayr} or, likely, by implementing some pertinently adapted version
of the finite-temperature density matrix renormalization group approach to quantum
spin chains \cite{feigwhite,karrasch}.

Alternatively, one may not be required to give up 
using fermionic coordinates, by  employing a systematic
renormalization group (RG) procedure to treat the effects of the 
bulk interaction on the scattering amplitudes at the junction. 
Among the various possible ways of implementing RG for 
junctions of interacting QWs, the two  most effective (and widely 
used) ones are certainly the poor man's fermionic renormalization
group (FRG)-approach, based upon a systematic 
summation of the “leading-log” divergences of the $S$-matrix
elements  at the Fermi momentum and typically yielding 
equations that can be analytically treated \cite{matgla1,matgla2,lal_1,nazarov_glazman}, and 
the functional renormalization group (fRG)-approach, based on the functional renormalization 
group method, leading to a set of coupled differential equations for the 
vertex part that, typically, can only be numerically treated \cite{meden1,meden2,meden3,meden4,meden5,meden5.1,meden6,meden7,medendue,medentre}. 
Both approaches are expected to apply only for a  
sufficiently weak electronic interaction in the quantum wires and, in 
this sense, they are  less general than the TLL-approach, which applies even
for a strong bulk interaction. Nevertheless, at  variance with the TLL-approach,
a RG-approach based on the use of fermionic coordinates leads to equations for the 
$S$-matrix elements valid at any scale and, thus, it 
allows for recovering the full scaling of  the conductance, all the way down to the infrared cutoff.

Aside from the remarkable merit of providing analytically tractable RG-flow equations, 
the FRG-approach, when applied to interacting spinful electrons, 
also accounts for the backscattering bulk interaction, which is 
usually neglected in the TLL-framework \cite{matgla1,matgla2}. Moreover, it can be 
readily generalized to describe junctions involving superconducting contacts, 
at the price of doubling the set of degrees of freedom, to treat particle-
and hole-excitations on the same footing \cite{belzig}. Nevertheless, 
FRG is known to suffer of limitations, when used to describe the crossover 
towards ''healed'' fixed points, where quantum impurities are renormalized 
towards boundary conditions corresponding to perfect conduction properties.
In particular, this leads to a scaling equation for the conductance which, in the pertinent 
asymptotic regime, is  not consistent with the one obtained from the bosonization 
approach. Using the fRG-approach allows for taking care of this flaw, as the fRG-technique 
typically takes into account the mutual feedback  from all the single-fermion 
scattering channels, and not only those from scattering processes between
different Fermi points (see Ref. [\onlinecite{meden6}] for a detailed discussion of
this point and for a careful comparison between FRG- and fRG-approaches). 
In fact,  fRG appears to be generically more accurate than FRG
(which can be in fact recovered from fRG under suitable approximations 
\cite{meden6}). Both approaches suffer, however, of the limitation on the 
bulk interaction, which must be weak, in order for the technique to be 
reliably applicable. 

To overcome such a limitation, in this paper, we study a junction of 
spinful interacting QWs by making  a combined use of 
bosonization and fermionization, that is, we go back and forth from fermionic to bosonic 
coordinates, and vice versa, to build ''dual-fermion'' representations of the junction in strongly interacting 
regimes. In  resorting from a bosonic to a fermionic problem, 
our approach is reminiscent of the refermionization scheme  used in  Ref. [\onlinecite{shulz_PRB}]  
to discuss  the large-distance behavior of the classical sine-Gordon model at 
the commensurate-incommensurate phase transition. Specifically, in \cite{shulz_PRB} 
the refermionization allows for singling out at criticality the low-energy two-fermion 
excitations from the one-fermion ones and to prove that the latter ones keep gapped 
along the phase transitions and do not contribute to the large-distance scaling of 
the correlations. At variance, in our case it is the second  of a two-step process, 
that ends up again into a fermionic ''dual fermionic'' model for the strongly-interacting system.  
The guideline to construct the appropriate novel fermionic  degrees of freedom is 
to eventually rewrite the relevant boundary interactions as bilinear functionals of the 
fermionic fields.   Specifically, moving  from the original fermionic coordinates to the 
TLL-bosonic description of the junction, we are able to warp from the weakly interacting regime to different 
strongly interacting regimes. Therefore, at appropriate values of the interaction-dependent
Luttinger parameters, we move back from the bosonic- to pertinent dual-fermionic 
coordinates, chosen so that  the relevant boundary interactions are bilinear functionals of the 
fermionic fields. Our mapping between dual coordinates is actually preliminary to the implementation 
of the RG-approach. Therefore, in principle, it could be equally well applied to extend both 
FRG and fRG to the strongly-interacting regime. Since, however, fRG usually requires resorting 
back to lattice models to implement a systematic numerical treatment of the flow equations 
for the interaction vertices, which goes beyond the scope of this work  
\cite{meden1,meden2,meden3,meden4,meden5,meden5.1,meden6}, we rather prefer to 
complement our dual mappings with   a  pertinently adapted version of the FRG-approach. 
 
The RG-approach formulated in fermionic coordinates, such as FRG, suffers of the limitation 
that it requires that relevant scattering processes
at the junction are fully encoded in terms of a single-particle $S$-matrix. 
While this is certainly the case at weak bulk interaction, a strong attractive 
interaction in either charge-, or spin-channel (or in both) is known to 
stabilize phases (RG attractive fixed points) at which two-particle 
scattering is the most relevant process at the junction 
\cite{furusaki,kane_fisher1,kane_fisher2,oca1,oca2}. Just because of the way it is formulated, 
the FRG-approach fails to describe many-particle scattering processes, 
even after improvements of the technique that allow to circumvent the 
constraint of having a small bulk interaction \cite{aristov_1,aristov_2,aristov_N}.
Resorting to the appropriate dual-fermion basis allows us to describe
within the FRG-approach also fixed points stabilized by 
many-particle scattering processes, as well as   fixed points whose properties 
have not   been mapped out within the TLL-framework in terms of a rotation matrix
such as, for instance, the mysterious-fixed point in the three-wire junction 
of spinless quantum wires 
studied in Ref. [\onlinecite{oca2}] and its counterpart in the junction of 
spinful quantum wires.
Moreover, in computing the conductance tensor along the RG-trajectories connecting 
fixed points of the phase diagram, we show how our approach, while being
consistent with the TLL-approach in the range of parameters where both of them apply, 
on the other hand allows for complementing the results of Refs. [\onlinecite{furusaki,kane_fisher1,
kane_fisher2,chamon}] about the two-wire and the three-wire junction, with a number of  additional 
results about the topology of their phase diagram and their conductance properties.

The paper is organized as follows:

\begin{itemize}

\item In Sec. \ref{two_wire} we apply our duality-complemented FRG-approach to a junction 
 of two interacting spinful quantum wire, discussing the results and comparing 
 them to those obtained within the bosonization approach 
 \cite{furusaki,kane_fisher1,kane_fisher2};
 
\item In Sec. \ref{three_wire} we apply our approach to a junction 
 of three interacting spinful quantum wire. We first compare our results 
 with those obtained within the bosonization approach \cite{chamon} and, 
 therefore, we show how our technique allows 
 for mapping out the full crossover of the conductance at the junction even 
 in strongly-interacting regions, typically not accessible with a fermionic  
approach;

\item We summarize our results and discuss possible further developments of 
our research in Sec. \ref{conclusions}, dedicated to the concluding remarks 
of our paper;

\item In the various appendixes we review mathematical techniques that 
are crucial for our derivation. Specifically, in Appendix \ref{frg_derivation},
we review the derivation of the FRG-equations 
for the $S$-matrix, in Appendix   \ref{boso_analysis}, we
review the basic bosonization rules for interacting one-dimensional quantum wires;
in Appendix \ref{ltr}, we provide some basic elements of linear transport theory for 
junctions of one-dimensional quantum wires. 
 
\end{itemize}

\section{Dual Fermionic variables and  renormalization group approach to the calculation of the conductance 
at a junction of two spinful interacting quantum wires}
\label{two_wire}
 
\begin{figure}
\includegraphics*[width=.55\linewidth]{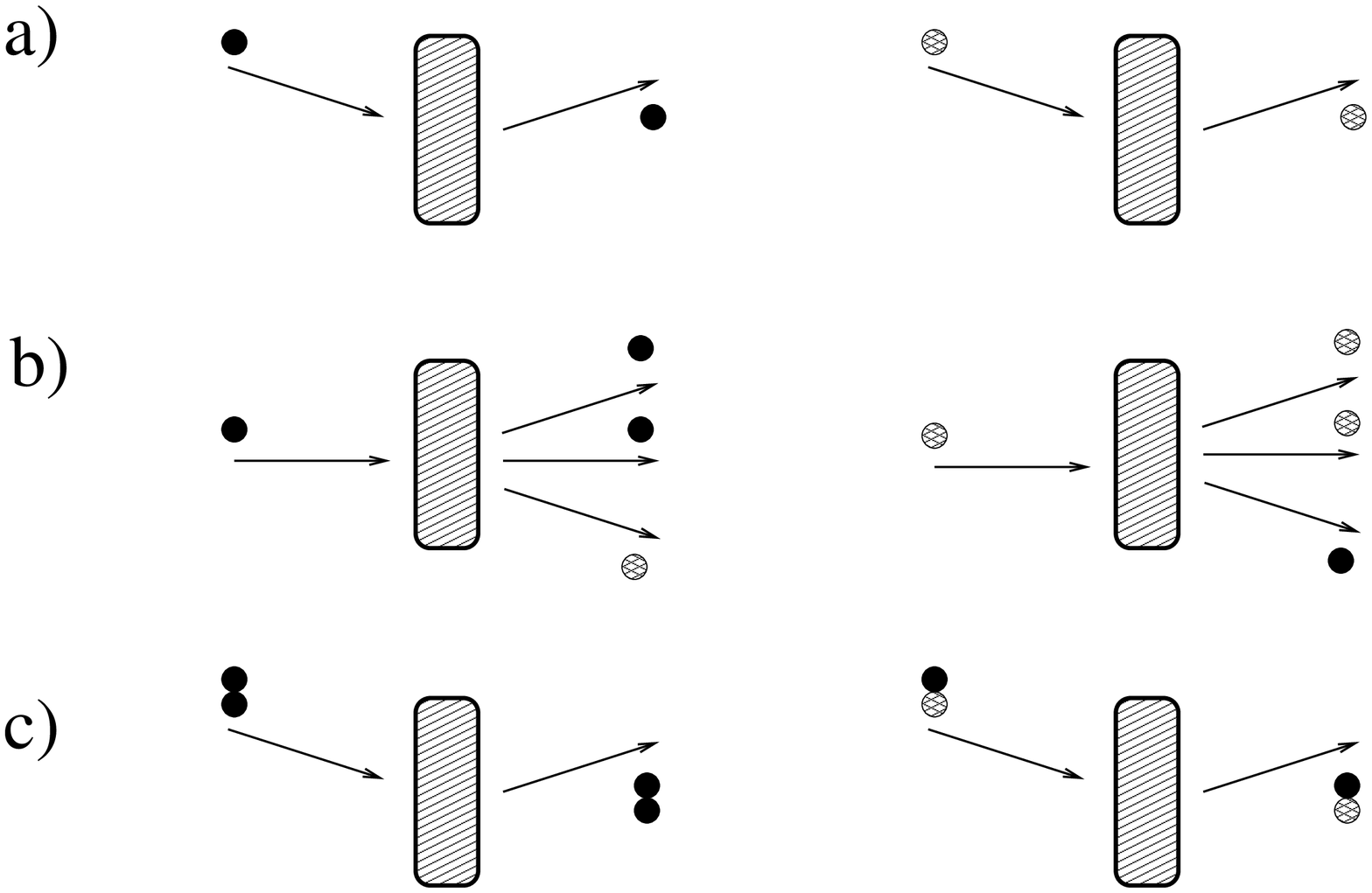}
\caption{Sketch of possible scattering processes at a two-wire junction (note
that incoming particles from wire $j$ can either be scattered into 
the same wire, or into a different wire): \\
{\bf (a)} Single-particle/single-hole  scattering processes. These 
are determined by $H_{\rm tun}$ in Eq. (\ref{th.1}) and are fully
described in terms of a single-particle $S$-matrix; \\
{\bf (b)} Many-body scattering processes in which one particle/one hole is 
scattered into two particles and one hole/two holes and one particle. These 
processes can be induced by boundary interaction Hamiltonians such as 
those in Eq. (\ref{abi.1}) and their proliferation requires resorting 
to a bosonic Luttinger-liquid description of the junction; \\
{\bf (c)} Scattering processes for a particle-particle and for 
a particle-hole pair.
These are again determined by the Hamiltonians in Eq. (\ref{abi.1}) and are the only
allowed processes in the presence of a strong repulsive (attractive)  interaction in 
the spin (charge) channel, and vice versa. On pertinently defining 
new fermionic coordinates, they can still be described in terms of a 
''single-pair'' $S$-matrix. } \label{scattering}
\end{figure}

To introduce and check the validity of our approach, in this section we discuss
a junction of two interacting spinful quantum wires. This appears to be quite an
appropriate place to test our technique: indeed, the two-wire junction has widely been 
studied in the past,  
both within the bosonization approach \cite{furusaki,kane_fisher1,kane_fisher2}, and by means of  standard
RG techniques for a weak bulk interaction, either using the FRG-approach \cite{matgla1,matgla2,lal_1,nazarov_glazman},
or the fRG-method\cite{meden1,meden2,meden3,meden4,meden5,meden5.1,meden6}. 
 The two-wire junction of spinful quantum wires is  described by 
the ($K=2$)  bulk Hamiltonian $H_{\rm Bulk} = H_0  + H_{\rm int}$, 
with

\begin{equation}
H_0 = 
- i v \sum_{ j = 1}^K \: \sum_{\sigma}\: 
\int_0^L \: dx\: 
\left\{ \psi_{R,j,\sigma}^{\dagger} ( x ) \partial_x 
\psi_{R,j,\sigma} ( x ) -\psi_{L,j,\sigma}^{\dagger} ( x ) 
\partial_x \psi_{L,j,\sigma} ( x ) \right\} 
\:\:\:\: , 
\label{fer.1}
\end{equation}
\noindent
with $L$ being the wire length (eventually sent to infinity at
the end of the calculations)
and the interaction Hamiltonian given by

\begin{eqnarray}
H_{\rm int} & \approx &\sum_{ j = 1}^K \: 
 \int_0^L \:  dx \: \biggl[ g_{j,1 , \parallel}
\psi_{R,j,\uparrow}^{\dagger} ( x ) \psi_{L,j,\uparrow}^{\dagger} ( x ) 
\psi_{R,j,\uparrow} ( x ) \psi_{L,j,\uparrow} ( x ) 
+g_{j,1 , \parallel} \psi_{R,j,\downarrow}^{\dagger} ( x ) 
\psi_{L,j,\downarrow}^{\dagger} (x ) \psi_{R,j,\downarrow} ( x ) 
\psi_{L,j,\downarrow} ( x ) \nonumber \\
 & + & g_{j,1 , \perp} \psi_{R,j,\uparrow}^{\dagger} ( x ) 
 \psi_{L,j,\downarrow}^{\dagger} ( x ) \psi_{R,j,\downarrow} ( x ) 
 \psi_{L,j,\uparrow} ( x ) +g_{j,1 , \perp}
 \psi_{R,j,\downarrow}^{\dagger} ( x ) \psi_{L,j,\uparrow}^{\dagger} ( x ) 
 \psi_{R,j,\uparrow} ( x ) \psi_{L,j,\downarrow} ( x ) \biggr]\nonumber \\
 & + & \sum_{ j = 1}^K \: 
 \int_0^L  \:  dx \: \biggl[g_{j,2 , \parallel}
\psi_{R,j,\uparrow}^{\dagger} ( x ) \psi_{L,j,\uparrow}^{\dagger} ( x ) 
\psi_{L,j,\uparrow} ( x ) \psi_{R,j,\uparrow} ( x ) +
g_{j,2 , \parallel}\psi_{R,j,\downarrow}^{\dagger} ( x ) 
\psi_{L,j,\downarrow}^{\dagger} ( x ) 
\psi_{L,j,\downarrow} ( x ) \psi_{R,j,\downarrow} (  x ) \nonumber \\
 & + & 
 g_{j,2 , \perp}\psi_{R,j,\uparrow}^{\dagger} ( x ) 
 \psi_{L,j,\downarrow}^{\dagger} ( x ) \psi_{L,j,\downarrow} ( x ) 
 \psi_{R,j,\uparrow} ( x ) +g_{j,2 , \perp}
 \psi_{R,j,\downarrow}^{\dagger} ( x ) 
 \psi_{L,j,\uparrow}^{\dagger} ( x ) \psi_{L,j,\uparrow} ( x ) 
 \psi_{R,j,\downarrow} ( x ) \biggr]
 \:\:\:\: . 
\label{eq:fermionic1}
\end{eqnarray}
\noindent
The various interaction strengths appearing in Eq. (\ref{eq:fermionic1}) 
are defined as 

\begin{eqnarray}
 g_{j,1 , \parallel}&=& V_{j,\uparrow\uparrow}(2k_{F})=
 V_{j,\downarrow\downarrow}(2k_{F}) \nonumber \\
 g_{j,2 , \parallel}&=& V_{j,\uparrow\uparrow}(0)=V_{j,\downarrow\downarrow}(0)
 \nonumber \\
 g_{j,1 , \perp}&=& V_{j,\uparrow\downarrow}(2k_{F})=V_{j,\downarrow\uparrow}(2k_{F})
 \nonumber \\
 g_{j,2 , \perp}&=& V_{j,\uparrow\downarrow}(0)=V_{j,\downarrow\uparrow}(0)
 \:\:\:\: , 
 \label{fer.7}
\end{eqnarray}
\noindent
with $ V_{\sigma , \sigma'} ( k )$ being the Fourier modes of 
the two-body interaction ''bulk'' interaction potential in 
the quantum wires (see Appendix \ref{frg_derivation} for the derivation and discussion of 
Eqs. (\ref{eq:fermionic1},\ref{fer.7}).)
 For a weak bulk interaction,  the 
most relevant contribution to $H_B$ is given by a linear
combination of the   operators 
$B_{(j , j') , \sigma , (X , Y ) }( 0 )$, defined as 

\beq
B_{(j , j') , \sigma , (X , X' ) }( 0 ) 
= \psi_{ X , j , \sigma}^\dagger ( 0 ) \psi_{X' , j' , \sigma } ( 0 ) 
\:\:\:\: , 
\label{appe.bou1}
\eneq
\noindent
with $X , X' = L , R$. Assuming
equivalence between the two wires and a spin-symmetric and spin-conserving 
boundary interaction, $H_B$ can be generically written as

\beq
H_{B} = \sum_{ X,X' = L , R }\: \sum_\sigma\:  
\{  [ \tau_{X , X' } B_{(1 , 2), \sigma , (X , X' ) } (0)  + {\rm h.c.} ]  + \sum_{ j = 1 , 2 }
\:   \mu_{ X , X'}   B_{(j , j), \sigma , (X , X' ) } (0)   \} 
\:\:\:\: . 
\label{th.1}
\eneq
\noindent
In addition to the contributions reported in Eq. (\ref{th.1}), 
 terms that are quadratic (or of higher order) in 
the $B$'s  can in principle  arise along RG-procedure, even if they are not present 
in the  ''bare'' Hamiltonian. For instance, the simplest 
higher-order boundary interaction terms 
consistent with spin conservation at the junction, $H_{2,0}$ and  $H_{0,2}$, 
are given by \cite{kane_fisher1,kane_fisher2,furusaki}
 
\begin{eqnarray}
 H_{2,0} &=&  V_{2,0} \: \sum_{ X , X' = R , L } \: \{ B_{ (X , X' ) , \uparrow , (1,2)} ( 0 ) 
 B_{ (X' , X ) , \downarrow , (1,2)} ( 0 ) + {\rm h.c.} \} \nonumber \\
  H_{0,2} &=& V_{0,2}  \:  \sum_{ X , X' = R , L } \: 
  \{B_{ (X , X' ) , \uparrow , (1,2)} ( 0 ) 
 B_{ (X' , X ) , \downarrow , (2,1)} ( 0 ) + {\rm h.c.} \}
 \:\:\:\: .
 \label{abi.1}
\end{eqnarray}
\noindent 
As it can be  shown using the bosonization approach, for a weak bulk interaction, higher-order
operators such as those in Eqs. (\ref{abi.1}) are highly irrelevant operators and, accordingly  they 
are typically ignored and one uses for $H_B$ the formula in Eq. (\ref{th.1}).
Physically, this means that the relevant scattering processes at the 
junction consist only of one single particle/hole scattered into one single particle/hole, 
such as those drawn in Fig. \ref{scattering} (a). These processes are  fully described by the single-particle $S$-matrix,
for which the renormalization group equations can be fully recovered using the 
technique we review in Appendix \ref{frg_derivation}. The symmetry requirements listed
above imply that   the single-particle $S$-matrix takes the block-diagonal form  

\beq
S_{ ( j , \sigma ) ; (j' , \sigma' ) } (k)= 
\delta_{ \sigma , \sigma'} \: S_{ j , j'} ( k )  
\:\:\:\: , 
\label{2wj.1}
\eneq
\noindent
with the $S (k)$-matrix being given by 
 
\beq
S (k) = \left[ \begin{array}{cc}
            r_k & t_k \\ t_k & r_k 
           \end{array} \right]
           \:\:\:\: , 
           \label{2wj.2}
           \eneq
           \noindent
and $r_k$ and $t_k$ respectively 
corresponding to the amplitude for the particle to be backscattered in the same wire, 
or transmitted into the other wire. In the 
following we will pose no particular constraints on the $r_k$'s and the $t_k$'s, except 
that, near the Fermi points, they are quite flat functions of $k$, without
displaying   particular features,  such as a resonant
behavior: accordingly, we   assume that the amplitudes are all computed at 
the Fermi level and   drop the $_k$ label (this is a specific case of the general  assumptions on the behavior 
of the $S$-matrix elements near by the Fermi surface that we make in 
Appendix \ref{frg_derivation}). To write the RG-equations for the $S$-matrix  
elements, one needs the Friedel matrix ${F}$ \cite{lal_1} which, in this specific case, 
is given by 
 
\begin{equation}
{F} =\frac{1}{2}\left[ \begin{array}{cccc}
\beta r & 0 & 0 & 0\\
0 & \beta r & 0 & 0\\
0 & 0 & \beta r & 0\\
0 & 0 & 0 & \beta r
\end{array}\right]
\;\;\;\; , 
\label{a.6}
\end{equation}
\noindent
with $ \beta= \frac{1}{2 \pi v} \: \left(-g_{1\parallel}-g_{1\perp}+g_{2\parallel}\right)$.
Taking into account the symmetries of the $S$- and of the $F$-matrix, 
the RG-equations for the amplitudes $r , t$ are obtained in the form

\begin{eqnarray}
\frac{dr}{d \ell} &=& \frac{\beta}{2}\left(r-r\left|r\right|^{2}-r^{*}t^{2}\right)= \beta r\left|t\right|^{2}
\nonumber \\
 \frac{dt}{d \ell} &=& -\beta t\left|r\right|^{2}=-\beta(t-t\left|t\right|^{2})
 \:\:\:\: . 
 \label{a.9}
\end{eqnarray}
\noindent
Equations (\ref{a.9}) must be supplemented with the RG-equation for the running strength $\beta$, 
which is given by  

\begin{equation}
\frac{d \beta}{d \ell }= \frac{1}{( 2 \pi v )^2} \{ 
\left(g_{1 , \perp}\right)^{2}+2g_{1 , \perp}\left(g_{2 , \perp}-g_{2, \parallel}+g_{1, \parallel}
\right) \} 
\:\:\:\: . 
\label{a.7}
\end{equation}
\noindent
(See Eqs. (\ref{eq:fermionic1},\ref{fer.7}) for the definition of the bulk interaction 
strengths $g_{ j , 1 ( 2)  , \perp} , g_{j , 1 ( 2 )  ,  \parallel}$: here 
we drop the wire index $j$ as the interaction strengths are assumed to be 
the same in each wire.)
Equations (\ref{a.9}), together with Eq. (\ref{a.7}) and Eqs. (\ref{eq:fermionic4}) for 
the running interaction strengths, constitute a closed set of equations, whose 
solution yields the scaling functions $r ( D ) , t ( D )$. From the explicit formulas
for the running scattering amplitudes, one may readily compute the charge- and 
the spin-conductance tensors, using the formulas derived in Appendix \ref{junoninter}.
As a result, due to the 
symmetries of the $S$-matrix, the charge- and the spin-conductance tensor
are equal to each other and both given by

\beq
G_c ( D )  = G_s ( D ) = \left[ \begin{array}{cc}
                                 \frac{ e^2}{\pi } - G ( D ) & G ( D ) \\ G ( D ) & 
                                  \frac{e^2}{\pi} - G ( D )
                                \end{array} \right]
\;\;\;\;,
\label{cnd.n1}
\eneq
\noindent
with $G ( D )  = \frac{e^2}{ \pi} | t ( D ) |^2 $. An explicit analytical formula can 
be provided for $G ( D )$ in some simple cases such as, for instance, if  $g_{1\perp}$ is 
fine-tuned to 0.  In this case, as it arises from Eq. (\ref{a.7}), $\beta$ keeps constant
along the RG-trajectories and, therefore,  one may 
exactly integrate Eqs. (\ref{a.9}) for  $r ( D ) $ and $t ( D ) $, obtaining  

\begin{figure}
\includegraphics*[width=1.\linewidth]{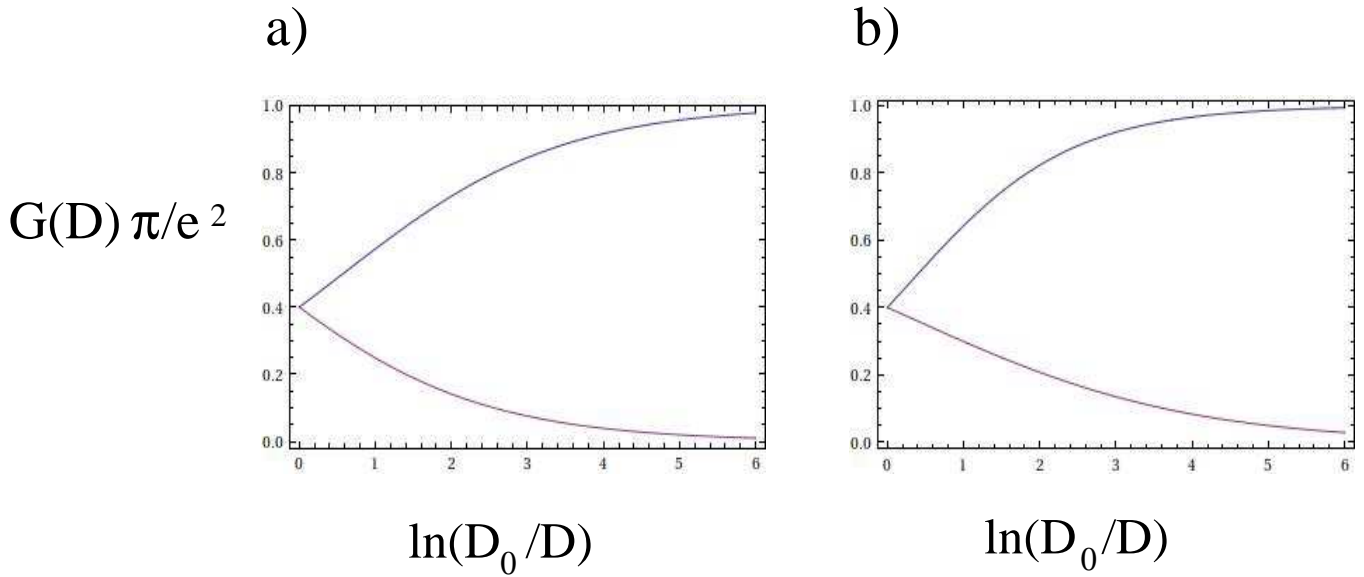}
\caption{{\bf (a)} Plot of $G ( D )$ {\it vs.} $\ln ( D_0 / D)$ as from 
Eq. (\ref{a.x1}) for $T_0 = 1 - R_0 = 0.4$, $\beta = 0.35$ (purple curve - corresponding to an 
effectively repulsive interaction), and  $\beta = -0.35$ (blue curve - corresponding to an 
effectively attractive interaction), with $T_0 = | t_0 |^2 , R_0 = | r_0 |^2$; \\
{\bf (b)} Plot of $G ( D )$ {\it vs.} $\ln ( D_0 / D)$ as given in 
Eq. (\ref{eq:t-epsilon-2})  for $T_0 = 1 - R_0 = 0.4$, $g_1 / ( 2 \pi v ) = 0.2$, $\gamma = 0.36$ (purple curve - corresponding to an 
effectively repulsive interaction), and  $\beta = -0.36$ (blue curve - corresponding to an 
effectively attractive interaction).} \label{plots_1}
\end{figure}

\begin{figure}
\includegraphics*[width=1.\linewidth]{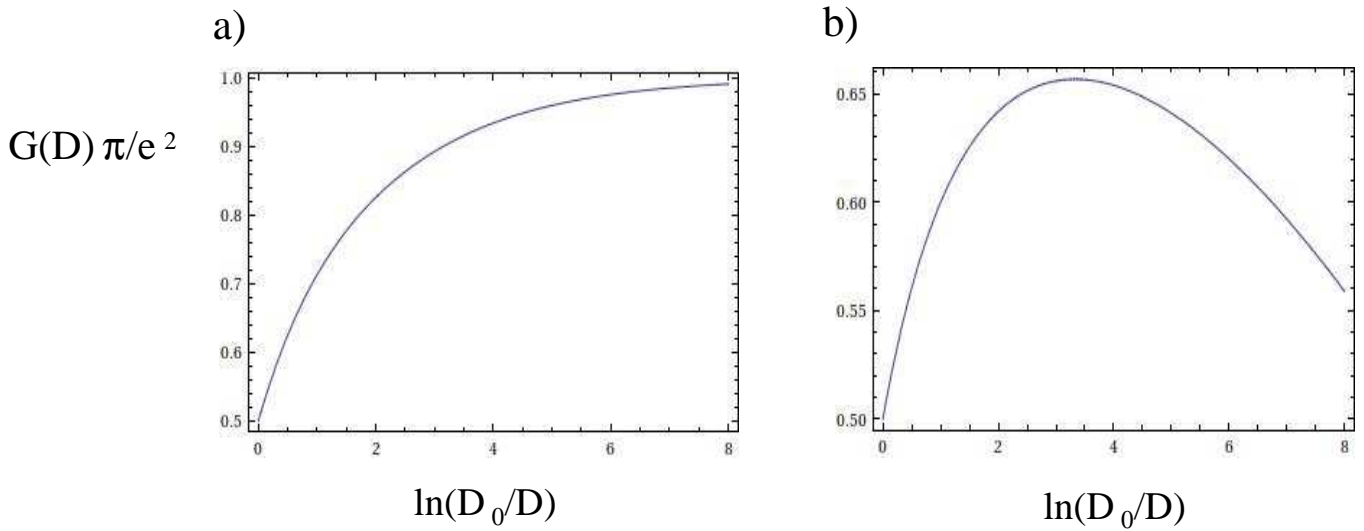}
\caption{{\bf (a)} Plot of $G ( D )$ {\it vs.} $\ln ( D_0 / D)$, obtained by 
numerically integrating  Eqs. (\ref{a.9},\ref{a.7},\ref{eq:fermionic4})
for $g_{1 , \parallel} (D_0 )  /(2 \pi v ) = g_{1 , \perp} (D_0 ) /(2 \pi v ) 
= g_{2 , \perp} (D_0 )  / ( 2 \pi v )  = 0.3$, $g_{2 , \parallel} (D_0 ) /(2 \pi v ) = 0$, $R_0 = T_0 = 0.5$; 
 \\
{\bf (b)} Same as in panel ${\bf (a)}$, but with $g_{2 , \parallel} (0)/(2 \pi v ) = 0.3$, as 
well.} \label{full_plots_2}
\end{figure}

\noindent

\begin{eqnarray}
 r ( D ) &=& \frac{t_0 \left|D /D_{0}\right|^{-\beta}}{\sqrt{\left|r_0 \right|^{2}+
\left|t_0 \right|^{2}\left|D /D_{0}\right|^{- 2\beta}}}
\nonumber \\
t( D )&=&
\frac{t_0 \left|D /D_{0}\right|^{\beta}}{\sqrt{\left|r_0 \right|^{2}+
\left|t_0 \right|^{2}\left|D /D_{0}\right|^{2\beta}}}
\:\:\:\: , 
\label{a.x1}
\end{eqnarray}
\noindent
with $r_0 , t_0$ corresponding to the ''bare'' scattering amplitudes
in Eq. (\ref{2wj.2}). 
Another case in which an explicit analytical solution
can be provided corresponds to having
 $g_{1\perp}=g_{1\parallel}=g_{1}$ and $g_{2\perp}=g_{2\parallel}=g_{2}$. In 
 this case, the set of Eqs. (\ref{eq:fermionic4}) collapse onto a set 
of two equations for $g_1 ( D ) , g_2 ( D ) $ which can be readily integrated,
yielding  the running interaction strengths

\begin{eqnarray}
g_{1}(D) &=& \frac{g_{1}}{1+ \frac{ g_{1} }{ \pi v} \ln\frac{D_{0}}{D} } \nonumber \\
g_{2}(D) &=& g_{2}-\frac{g_{1}}{2}+\frac{1}{2}\frac{g_{1}}{1+\frac{ g_{1} }{ \pi v } \ln\frac{D_{0}}{D}}
\:\:\:\: ,
\label{rint.strengths}
\end{eqnarray}
\noindent
and $\beta ( D ) =  [ g_2 ( D ) - 2 g_1 ( D ) ] / ( 2 \pi v ) $. Once $\beta ( D )$ is known, 
Eqs. (\ref{a.9}) can be integrated, yielding 

\beq
G ( D )  = \frac{e^2}{\pi}\: \left[ \frac{T_0
\left[1+ \frac{ g_{1} }{ \pi v } \ln\left|\frac{D_{0}}{D}\right|\right]^{3/2}\left|D/D_{0}\right|^{2 \gamma}}{
R_0 + T_0 \left[1+ \frac{ g_{1} }{ \pi v} \ln\left|\frac{D_{0}}{D}\right|
\right]^{3/2}\left|D/D_{0}\right|^{2\gamma}} \right]  \;\;\;\; ,
\label{eq:t-epsilon-2}
\eneq
\noindent
with $\gamma=\left(-\frac{g_{1}}{2}+g_{2}\right) / ( 2 \pi v ) $, $T_0 = | t_0 |^2 , R_0 = | r_0 |^2$. 
As an example of typical scaling plots for $G ( D )$ in the simple cases discussed 
before, in Fig. \ref{plots_1} we plot $G(D) \pi / e^2$ {\it vs.} $\ln ( D_0 / D )$, as 
from  Eq. (\ref{a.x1}) (panel {\bf (a)}) and from Eq. (\ref{eq:t-epsilon-2}) 
(panel {\bf (b)}), with the values of the parameters reported in the caption.
Consistently with the results obtained within Luttinger liquid framework \cite{furusaki,kane_fisher1,
kane_fisher2}, 
$ G ( D ) \pi / e^2$ either flows to 0 for an effectively repulsive interaction ($\beta , \gamma > 0$), or
to 2 (the maximum value allowed by unitarity), for an effectively attractive 
interaction ($\beta , \gamma < 0$). For general values of the interaction
strengths, the equations have to be numerically integrated. 
In Fig. \ref{full_plots_2}, we provide some examples of 
scaling of $G ( D )$ {\it vs.}  $\ln ( D_0 / D )$ in the general case. It is 
important to stress \cite{lal_1} that, due to the nontrivial renormalization
group flow of the interaction strengths, the flow of $G ( D)$ can be a nonmonotonic
function of $D$ for some specific values of the interaction strengths. It would be
interesting to check such a feature in a real life two-wire junction: remarkably, 
this prediction is only obtained within the FRG-approach, in
which it is possible to account for the flow of the running interaction strengths, as 
well.

\noindent

\noindent
The possibility of mapping out the full crossover of the conductance as a function 
of the scale $D$ is possibly the most important feature of the FRG approach. Yet, since, as 
we discuss to some extent in Appendix \ref{frg_derivation}, the validity of the FRG technique is grounded on 
the assumption that all the relevant scattering processes at the 
junction are described by the single-particle $S$-matrix \cite{matgla1,matgla2,lal_1,nazarov_glazman}, it 
breaks down when attempting to  recover the full crossover of the conductance towards
fixed points where multi-particle scattering is the most relevant process 
at the junction, such as the strongly coupled fixed point stabilized by either $H_{2,0}$
or $H_{0,2}$ in  Eq. (\ref{abi.1}) \cite{furusaki,kane_fisher1,kane_fisher2}. Technically, what 
happens is that, as soon as boundary operators such as $H_{2,0}$
or $H_{0,2}$   become relevant, the proliferation of low-energy many-body
scattering processes such as the one we sketch in Fig. \ref{scattering} {\bf (b)} 
invalidates the single-particle $S$-matrix description of the junction dynamics. 
 Nevertheless, some many-body scattering processes can be strongly limited by having, 
for instance, a strong repulsive interaction among particles with the 
same spin and a strong attractive interaction among particles with the 
same charge. In the bosonic framework, this corresponds to having values of  the Luttinger parameters
in Eqs. (\ref{an.14})  such that $g_c \geq 2 , g_s \ll 1$. Indeed, in this limit on   one hand, 
the strong spin repulsive 
interaction forbids the single-particle processes described by $H_B$ in Eq. (\ref{th.1}) (at small 
values of the boundary coupling strengths $\tau_{ X , X'} , \mu_{ X , X'}$ this can be 
readily seen from  the explicit result for the scaling dimension of $H_{B}$ computed within 
the bosonization approach, which is $x_{B , weak} = 1 - \frac{1}{2 g_c} - \frac{1}{2 g_s}$, 
which becomes $\gg 1$, corresponding to a largely irrelevant operator). On the other hand, 
one expects that the strong charge attraction stabilizes tunneling of composite 
objects carrying zero spin, such as two-particle pairs, as the one 
depicted at the left-hand panel of Fig. \ref{scattering} {\bf (c)}. As a consequence, due to 
the fact that these are 
again one-into-one scattering processes, one expects that 
 it is possible to choose the effective low-energy degrees of freedom
of the system to resort to a single-particle $S$-matrix in the new coordinates. 
In fact, this is the idea behind the dual fermion approach we are
going to discuss next. When resorting to dual fermion coordinates, 
an important issue  is related to whether the vacuum states at a fixed 
particle number \cite{von_delft} for the original and the dual-fermions are the same.
In fact, while dual fermion formalism only captures “composite” excitations e.g. two-particle 
 states in the $g_c \sim 2, g_s \ll 1$-regime, at such values of 
 the parameters,  these states are the only ones 
that at low energy are effectively able to tunnel across the junction (that is, the 
only ones whose tunneling is described by a non-irrelevant operator). So, as long 
as one is only concerned about states relevant for low-energy tunneling across the 
junction (that is, states relevant for the calculation of the dc-conductance tensor 
of the junction), one can effectively assume that the fixed-particle number vacuum 
states are the same in terms of the new (dual) and of the old fermions. 
The definition of the dual fermion operators strongly depends on the boundary
conditions of the various fields at the junction. Accordingly, in the following we
define different dual fermion operators in different regimes of values of the boundary 
interaction, and eventually show that, whenever two different sets of dual coordinates 
apply to the same region, they yield the same results, as they are expected to. 

Let us begin with the weak boundary interaction regime. Referring to the bosonization 
formulas of Appendix \ref{boso_analysis}, this corresponds to assuming Neumann (Dirichlet) 
boundary conditions for all the $\Phi$ ($\Theta$)-fields in Appendix \ref{boso_analysis} and, in addition, 
to equating the Klein factors so that $\eta_{ R , \sigma , j } = \eta_{R , \sigma , j}$, 
$\forall \sigma , j$. In this limit, one may respectively rewrite  $H_{0 , 2}$  and 
$H_{2 , 0}$ in bosonic coordinates as 

\begin{eqnarray}
 H_{ 2 , 0 } &=&  v _{2,0} \cos [ \Phi_{1,c} ( 0  ) - \Phi_{2 , c } ( 0 ) ] \nonumber \\
 H_{ 0 , 2 } &=&  v _{0,2} \cos [ \Phi_{1,s} ( 0  ) - \Phi_{2 , s } ( 0 ) ]
\:\:\:\: ,
\label{next.1}
\end{eqnarray}
\noindent
with $v_{2,0} \propto V_{2,0} , v_{0,2} \propto V_{0,2}$. 
The scaling dimensions of the operators in Eqs. (\ref{next.1}) are respectively 
given by $x_{2,0} = 2 / g_c , x_{0,2} = 2 / g_s$. Thus, in 
the regime of a strongly attractive  interaction in the charge (spin)-channel and 
strongly repulsive interaction in the spin (charge)-channel, $H_{ 2 , 0 } $ ($H_{0,2}$) 
may become the most relevant boundary operator at weak boundary coupling. 
The strategy of our dual fermion approach consists in defining 
a novel set of fermionic fields, in terms of which the operators in Eqs. (\ref{next.1}) 
are realized as bilinears, similar to the $B$-operators in Eq. (\ref{appe.bou1}). 
To be specific,  let us 
introduce the center-of-mass and the relative fields in the charge- and in the spin-sector, 
respectively given by

\begin{eqnarray}
 \Phi_{c (s)}  ( x ) &=& \frac{1}{\sqrt{2}} [ \Phi_{ 1 , c (s)} ( x ) + \Phi_{ 2, c (s)} ( x )] \nonumber \\
  \Theta_{c (s)}  ( x ) &=& \frac{1}{\sqrt{2}} [ \Theta_{ 1 , c (s)} ( x ) + \Theta_{ 2, c (s)} ( x )]
  \:\:\:\: , 
  \label{nn.a1}
\end{eqnarray}
\noindent
and

\begin{eqnarray}
 \varphi_{c (s)} ( x ) &=& \frac{1}{\sqrt{2}} [ \Phi_{ 1 , c (s)} ( x ) - \Phi_{ 2, c (s)} ( x )] \nonumber \\
 \vartheta_{c (s) } ( x ) &=& \frac{1}{\sqrt{2}} [ \Theta_{ 1 , c (s)} ( x ) - \Theta_{ 2, c (s)} ( x )]
  \:\:\:\: . 
  \label{nn.a2}
\end{eqnarray}
\noindent
Next, let us perform the canonical transformation to a new set of 
bosonic fields, defined as 

\beq
\left[ \begin{array}{c}
        \bar{\Phi}_{ c(s)} ( x ) \\ \bar{\Theta}_{ c (s)} ( x ) \\
        \bar{\varphi}_{c(s)} ( x ) \\ \bar{\vartheta}_{ c ( s ) } ( x  ) 
       \end{array} \right] 
       = 
       \left[ \begin{array}{cccc}
               \sqrt{2} & 0 & 0 & 0 \\ 0 & \frac{1}{\sqrt{2}} & 0 & 0 \\ 0 & 0 & \sqrt{2} & 0 \\ 
               0 & 0 & 0 & \frac{1}{\sqrt{2}}
              \end{array} \right]\left[ \begin{array}{c}
        \Phi_{ c(s)} ( x ) \\ \Theta_{ c (s)} ( x ) \\
        \varphi_{c(s)} ( x ) \\ \vartheta_{ c ( s ) } ( x  ) 
       \end{array} \right] 
        \:\:\:\: . 
        \label{nn.a3}
        \eneq
        \noindent
It is worth stressing that the transformation in Eqs. (\ref{nn.a3}) relate to each other bosonic operators 
at a given position in real space. Since the correspondence rules between the bosonic  and 
the (original or dual) fermionic fields, summarized in Appendix \ref{boso_analysis}, are local in real space, 
as well, one concludes that, written in terms of dual fermionic coordinates,  the boundary 
interaction Hamiltonian $H_B$ is still local and that the dynamics far from the junction 
can be fully encoded within dual fermion scattering states.
Now, assuming $g_s \ll 1$,  $g_c = 2  + \delta g_c$, with $ | \delta g_c | \ll 1$, we see that 
$g_s \ll 1$ makes $H_{0,2}$ strongly irrelevant. This fully suppresses spin transport across
the junction and, therefore,  we may just focus onto charge transport, ruled by 
$H_{2,0}$.  In fact, it appears that single-spinful particle-tunneling processes are already suppressed 
against two-particle pair tunneling processes as soon as $g_s < 2/3$. As conservation of 
spin symmetry implies $g_s = 1$, in order to realize the condition above   one may, for
instance, think of two coupled spinless interacting one-dimensional electronic systems 
(which could possibly realized as semiconducting quantum wires in the presence of spin-orbit
and Zeeman interactions), with a mismatch in the Fermi momenta that prevents the interaction 
from opening a gap in the fermion spectrum. The two channels can, therefore, be regarded as
the two opposite spin polarization, although without any symmetry implying $g_s = 1$. 
To rewrite this latter operator as a bilinear functional of fermionic operators,
 we define the spinless chiral fermionic fields
$\chi_{R,j} ( x ) , \chi_{L , j } ( x )$ as 

\begin{eqnarray}
 \chi_{R,j} ( x ) &=& \eta_{R , j } \; e^{ \frac{i}{2} [ \bar{\Phi}_{  c } ( x ) - 
 (-1)^j \bar{\varphi}_c ( x ) +  \bar{\Theta}_{ c } ( x ) -   (-1)^j \bar{\vartheta}_c ( x )
 ]} \nonumber \\
  \chi_{L,j} ( x ) &=& \eta_{L , j }  \; e^{ \frac{i}{2} [ \bar{\Phi}_{  c } ( x ) +
 (-1)^j \bar{\varphi}_c ( x ) +  \bar{\Theta}_{ c } ( x ) +   (-1)^j \bar{\vartheta}_c ( x )
 ]} 
    \;\;\;\; ,
  \label{resc.2}
\end{eqnarray}
\noindent
with $j = 1 , 2$ and with $\eta_{R , j } , \eta_{L , j }$ being real fermionic Klein factors. 
''Inverting'' the bosonization procedure outlined in Appendix \ref{boso_analysis} into 
a pertinent re-fermionization to spinless fermions, we find that the 
bulk Hamiltonian for the $\chi$-fermions is given by

\begin{eqnarray}
H_{\rm c; bulk } &=& - i u \sum_{ j = 1,2 } \;  \int_0^L  \; d x \; \{ \chi_{R , j }^\dagger ( x ) 
\partial_x \chi_{R , j } ( x ) - \chi_{L , j }^\dagger ( x )  \partial_x \chi_{L , j } ( x ) \} 
\nonumber \\ 
- &\frac{u  \pi \delta g_c}{2}& \sum_{ j = 1,2 } \;  \int_0^L  \; d x \; 
: \chi_{R , j }^\dagger ( x )  \chi_{R , j } ( x ) : : \chi_{L , j }^\dagger ( x )  \chi_{L , j } ( x ) : 
    \;\;\;\; ,
  \label{resc.3}
\end{eqnarray}
\noindent
with the velocity $u \propto v$. 
The $\chi_{ L / R , j }$-fields are the appropriate degrees of freedom to 
describe pair scattering at the junction in terms of a single-particle $S$-matrix. In 
order to prove that it is so, we note that $H_{2,0}$ can be regarded as the bosonic expression for the boundary
weak coupling limit of a tunnel Hamiltonian for the spinless fermions, $H_{\rm c , tun}$, 
given by
 
 \beq
 H_{\rm c , tun} = v_{2,0}  \{ \chi_{1 }^\dagger ( 0 )  \chi_{2 } ( 0 ) +  
 \chi_{2 }^\dagger ( 0 )  \chi_{1 } ( 0 ) \} 
 \;\;\;\; ,
\label{next.5}
\eneq
\noindent
with  $\chi_{j } ( 0 ) =  \chi_{R , j } ( 0 ) + \chi_{L , j } ( 0 )$. 
While the strong spin repulsion sets the spin conductance
tensor to 0, the charge conductance can nevertheless be different from zero, due 
to zero-spin pair-tunneling across the junction. Once the RG-flow for the $S$-matrix 
elements describing $\chi$-fermion scattering at the junction has been derived 
as we did before, using the formulas we report in Appendix \ref{junoninter} and 
the expression of the charge current operator in wire $j$ in terms of the dual fermionic fields:
\beq
J_{c , j } ( x ) = e u \sqrt{2} \: \{ : \chi_{ R , j }^\dagger ( x ) \chi_{ R , j } ( x ) : - 
: \chi_{ L , j }^\dagger ( x ) \chi_{ L , j } ( x ) : \}
\:\:\:\: , 
\label{next.6a}
\eneq
\noindent
 we  obtain that the charge conductance tensor scales according to

\beq
G_c ( D ) = \left[ \begin{array}{cc}
\frac{e^2}{\pi} - G ( D ) &   G ( D )    \\   G ( D )    &   \frac{e^2}{\pi} - G ( D )              
                   \end{array} \right]
\;\;\;\; , 
\label{next.6}
\eneq
\noindent
with 

\beq
G (D)  = \frac{e^2}{\pi} \: \left[ \frac{T_0 | D / D_0 |^{ -  \frac{ \delta g_c }{2} } }{
R_0 + T_0 | D / D_0 |^{ -   \frac{ \delta g_c }{2} } } \right]
\:\:\:\: , 
\label{re3.5}
\eneq
\noindent
and the bare reflection and transmission coefficients respectively given by

\begin{eqnarray}
 R_0 &=&  \left| \frac{u^2 - v_{2,0}^2}{u^2 + v_{2,0}^2} \right|^2 \nonumber \\
T_0 &=&  \left| \frac{ 2 u v_{2,0}}{u^2 + v_{2,0}^2} \right|^2 
\:\:\:\: .
\label{re3.6}
\end{eqnarray}
\noindent
\begin{figure}
\includegraphics*[width=.6\linewidth]{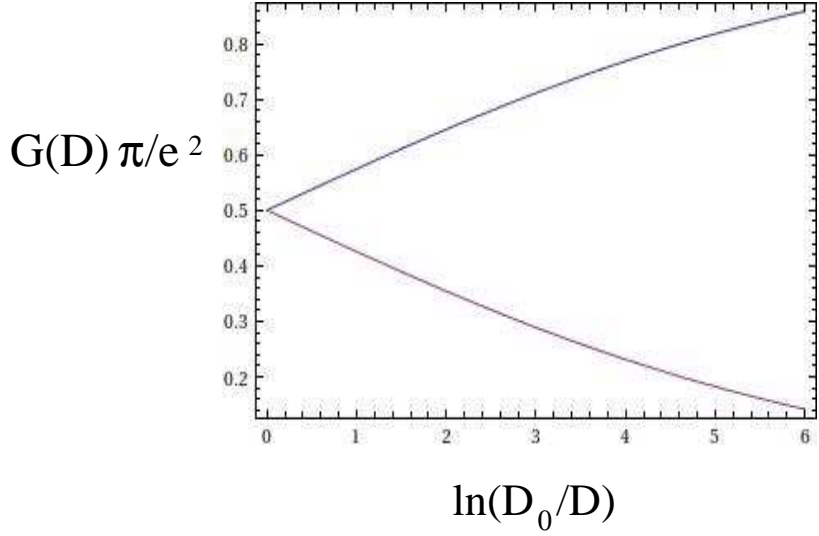}
\caption{Plot of $G_c ( D ) $ {\it vs.} $\ln ( D_0 / D )$ as from 
Eq. (\ref{re3.5}) for $T_0 = R_0 = 0.5$ and for $\delta g_c$ respectively 
equal to 0.6 (blue curve) and to -0.6 (purple curve).} \label{scal_interac_1}
\end{figure}
\noindent
In Fig. \ref{scal_interac_1}, we plot $G ( D )$ versus $\ln ( D_0 / D ) $ 
in two paradigmatic cases, respectively corresponding to 
$\delta g_c > 0$ and to $\delta g_c < 0$. To our knowledge, this is the
first example of a full scaling plot of the conductance for a junction 
of strongly interacting one-dimensional quantum wires. While, on one hand, this  shows the 
effectiveness of our approach in describing the crossover of 
the conductance towards the spin-insulating charge-conducting fixed point, on 
the other hand,  one has also to prove the consistency of an 
 effective theory strongly  relying on the weak boundary coupling assumption 
 with an RG-flow taking the system  all the way 
down to the perfectly charge-conducting fixed point, corresponding to 
the strongly interacting limit of the boundary interaction \cite{furusaki,kane_fisher1,kane_fisher2}.
When $\delta g_c > 0$, the relevance of $H_{2, 0 }$ 
drives the system towards the strongly boundary interaction limit in the charge channel,
corresponding to pinning $\varphi_c ( 0 )$ and, accordingly, to imposing 
Neumann boundary conditions on $\vartheta_c ( 0 )$. Since $\Phi_c ( 0 ) $ 
does not appear in the boundary interaction, one assumes that it still
obeys Neumann boundary conditions and, accordingly, that $\Theta_c ( 0 )$ is 
pinned at a constant value. We now prove that these boundary conditions are recovered 
by taking the strongly coupled limit of  $H_{\rm c , tun}$ in  Eq. (\ref{next.5})
and using the refermionization rules in Eqs. (\ref{resc.2}).  Indeed, on making 
the strong-coupling assumption, $\left| \frac{v_{ 2 , 0 }}{u} \right| \gg 1$,   as   from 
Eqs. (\ref{re3.6}), one obtains $ R_0 \to 0 , T_0 \to 1$, that is, the boundary 
conditions correspond to perfect transmission from wire-1 to wire-2, and vice versa. In terms of the 
dual fermionic fields, this corresponds to the conditions 

\begin{eqnarray}
 \chi_{R , 2 } ( x ) &=& e^{i \lambda} \: \chi_{L , 1 } ( - x ) \nonumber \\
  \chi_{R , 1 } ( x ) &=& e^{- i \lambda} \: \chi_{L , 2 } ( - x ) 
  \;\;\;\; ,
  \label{next.a1}
\end{eqnarray}
\noindent
with $\lambda$ being some nonuniversal phase. From  Eqs. (\ref{resc.2}), one 
sees that Eqs. (\ref{next.a1}) imply Dirichlet boundary conditions 
at $x = 0 $ for both $\bar{\varphi}_c ( x ) $ and $\bar{\Theta}_c ( x )$, with 
the dual fields $\bar{\vartheta}_c ( x ) , \bar{\Phi}_c ( x )$ obeying Neumann
boundary conditions.  This is exactly the same result one would obtain working in bosonic 
variables by sending to $\infty$ the interaction strength $v_{2,0}$ in Eq. (\ref{next.1}).
Due to the strong repulsion in the spin channel, such a fixed point   corresponds to perfect
transmission in the charge channel, but perfect reflection in the spin channel, that is, 
it must be identified with the non-symmetric charge-conducting spin-insulating 
phase of Refs. [\onlinecite{furusaki,kane_fisher1,kane_fisher2}].  
To conclude the consistency check, we note that, on alledging for 
additional backscattering contributions to $H_{ c , tun}$ of the generic form $  \mu_1  \chi_{R , 1}^\dagger ( 0 ) 
 \chi_{R , 1}^\dagger ( 0 )   + \mu_2 \chi_{L , 1}^\dagger ( 0 ) 
 \chi_{L , }^\dagger ( 0 )$ (which play no role at weak coupling) and using again 
 Eqs. (\ref{next.a1}), one obtains the bosonic operators 

\beq
\tilde{H}_{\rm c , tun} \sim  
\mu  \cos \left[ \bar{\vartheta}_c ( 0 ) 
\right]
\:\:\:\: , 
\label{abi.14}
\eneq
\noindent
with $\mu$ being some nonuniversal constant. Equation (\ref{abi.14}) corresponds to the bosonic version of the leading boundary
perturbation at the non-symmetric charge-conducting spin-insulating fixed point 
\cite{furusaki,kane_fisher1,kane_fisher2}. 

Our  approach also allows for analyzing the complementary situation in which 
$g_c \ll 1$ and $g_s \sim 2$. In this case, one expects that the strong repulsion in the charge
channel and the strong attraction in the spin channel stabilize single-pair tunneling 
processes at the junction such as those sketched at the right-hand panel of
Fig. \ref{scattering} {\bf (c)}, that is, tunneling of particle-hole pairs, with total
spin 1. Again, for $g_s = 2$, the $S$-matrix describes single-particle into single-particle
scattering processes, once it is written in the appropriate basis. To select the 
pertinent degrees of freedom, we therefore repeat the refermionization procedure 
in Eq. (\ref{resc.2}), by just exchanging the charge- and the spin-sector with each 
other. Of course, charge- and spin-conductance are exchanged with each other, compared to 
the previous situation and, accordingly, 
the flow will be towards the charge-insulating spin-conducting fixed point
of  Refs. [\onlinecite{furusaki,kane_fisher1,kane_fisher2}]. An important remark, however, concerns the 
effects of a possible residual interaction, which, as we did before, can be in 
principle introduced for accounting for $g_s$ slightly different from 2. Indeed, 
a term in the ''residual'' bulk interaction Hamiltonian such as the one 
$\propto g_{j , 1 , \perp}$ in Eq. (\ref{eq:fermionic1}), once expressed in 
terms of the fermionic fields in Eqs. (\ref{resc.2}) would take the form 

\beq
 H_\delta   = \sum_{ j  =1}^2 \: m_j \: \int \: d x \:  \{ \chi_{ R ,   j }^\dagger ( x )
 \chi_{ L ,   j } ( x ) + \chi_{ L ,  j}^\dagger ( x ) \chi_{ R ,   j } 
 ( x ) \}  
 \:\:\:\: ,
 \label{sp.4}
\eneq
\noindent
with $m_j \propto g_{j , 1 , \perp}$, which would open a bulk gap in 
the single-$\chi$ fermion spectrum, thus making the whole system behave as 
a bulk spin insulator. Therefore, in order to recover the correct physics 
of the charge-insulating spin-conducting fixed point, we must assume that all 
the $g_{j , 1 , \perp}$ are tuned to zero, which is typically the case 
when resorting to the bosonic approach to spinful electrons \cite{furusaki,kane_fisher1,kane_fisher2}.

As we have just shown, resorting to pertinent dual-fermion operators allows for mapping out
the full crossover with the appropriate energy scale of the charge and/or spin conductance
of a junction in region of values of the interaction parameters in which one is 
typically forbidden to use the standard weak-coupling formulation of either FRG, or fRG.
 In the following, we apply our  technique to the 
spinful three-wire junction studied using the bosonization approach in Ref. [\onlinecite{chamon}],
and  will recover the full crossover of the conductance tensor in regions typically not
accessible in the bosonic formalism.

 \section{Dual Fermionic variables and  renormalization group approach to the calculation of the conductance 
at a junction of three spinful interacting quantum wires}
 \label{three_wire}
 
\begin{figure}
\includegraphics*[width=.55\linewidth]{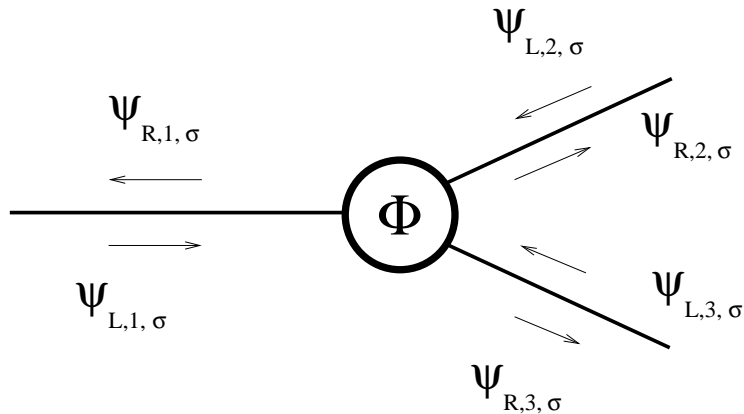}
\caption{Sketch of a three-wire junction of spinful quantum wires pierced by a magnetic 
flux $\Phi$.} \label{3wj}
\end{figure}

 We now  consider a three-spinful-wire junction, such as the one 
we sketch in Fig. \ref{3wj}. Resorting to the appropriate 
fermionic variables, we  generalize to strongly interacting regions
the weak-coupling FRG-approach. As a result,    we  
map out the full dependence  of the conductance tensor on  the low-energy
running cutoff  scale even in strongly-interacting regions of the 
parameter space. Eventually, we discuss the consistency of our results 
about the phase diagram of the junction with those obtained in 
Ref. [\onlinecite{chamon}], particularly showing how our technique can be used to recover
informations that typically cannot be derived within the bosonization approach used there.  
Consistently with Ref. [\onlinecite{chamon}], in the following, we  make the 
simplifying assumption that, in the weakly interacting regime, the bulk interaction is 
purely intra-wire and is the same in all the three wires. 
In fact, while this assumption is already expected to yields quite a rich phase 
diagram \cite{chamon}, in principle our approach can be readily 
generalized to cases of different bulk interactions 
in different wires, such as the one considered in Ref. [\onlinecite{chamon_2}].

\noindent

\subsection{The weakly interacting regime}
\label{wir}

\begin{figure}
\includegraphics*[width=.8\linewidth]{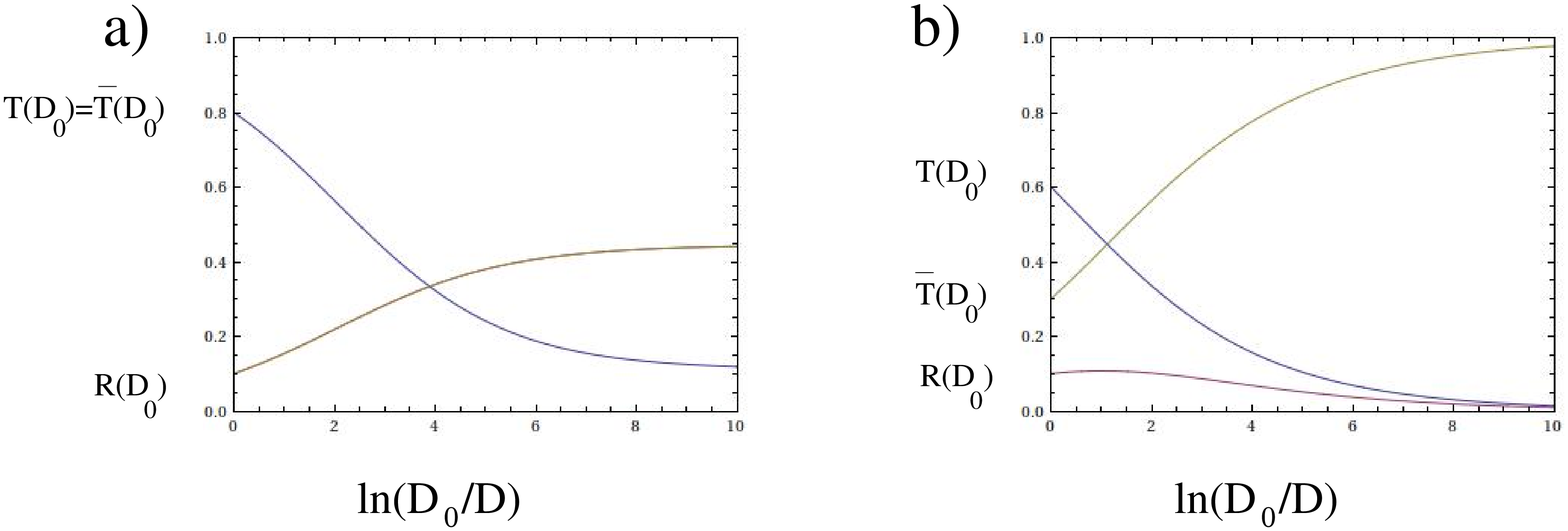}
\caption{Renormalization group flow of the scattering coefficients for a three-wire
junction for $g_{1 , \perp} = 0$, $\beta$ constant and equal to -0.3, and 
different choices of the initial values of the scattering coefficients: \\
{\bf (a)}   Renormalization group flow corresponding to  $T ( D_0 ) = \bar{T} ( D_0 ) = 0.1$, $R ( D_0 ) = 0.8$.
The curves corresponding to $T ( D )$ and to $\bar{T} ( D ) $ {\it vs} $\ln ( D_0 / D )$ 
collapse onto the single red curve of the graph, while the flow of   $R ( D ) $ {\it vs} $\ln ( D_0 / D )$
is described by the blue curve. As $D_0 / D$ grows, the scattering coefficients flow towards
the asymptotic values corresponding to the $M$-fixed point; \\
 {\bf (b)}   Renormalization group flow corresponding to   $T ( D_0 ) = 0.1 , \bar{T} ( D_0 ) = 0.3 , R ( D_0 ) = 0.6$.
As $D_0 / D$ grows, the scattering coefficients flow towards
the asymptotic values corresponding to the $\chi_{--}$-fixed point.} \label{plot_3j_1}
\end{figure}

\begin{figure}
\includegraphics*[width=.8\linewidth]{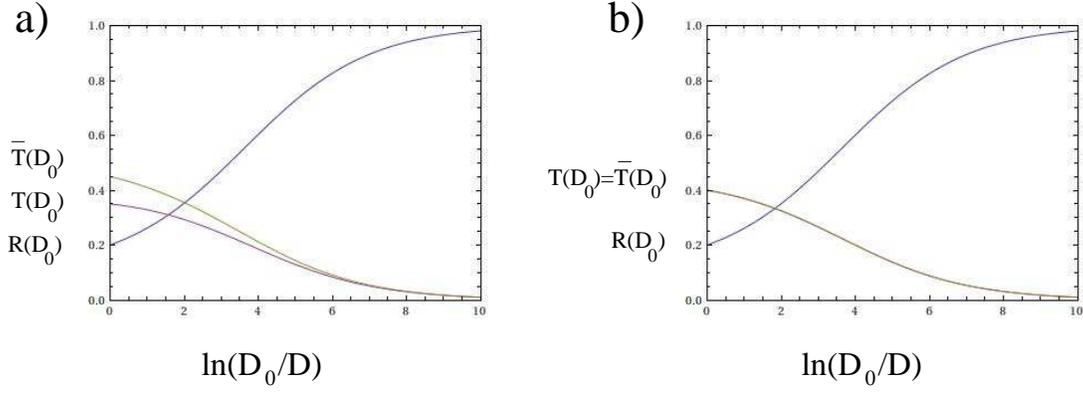}
\caption{Renormalization group flow of the scattering coefficients for a three-wire
junction for $g_{1 , \perp} = 0$, $\beta$ constant and equal to 0.3, and 
different choices of the initial values of the scattering coefficients: \\
{\bf (a)}   Renormalization group flow corresponding to  $T ( D_0 ) = 0.35, \bar{T} ( D_0 ) = 0.45 , R ( D_0 ) = 0.2$; \\
 {\bf (b)}   Renormalization group flow corresponding to   $T ( D_0 ) =  \bar{T} ( D_0 ) = 0.4 , R ( D_0 ) = 0.2$.
As $D_0 / D$ grows, in both cases the scattering coefficients flow towards
the $NN$-fixed point.} \label{plot_3j_2}
\end{figure}

\begin{figure}
\includegraphics*[width=.8\linewidth]{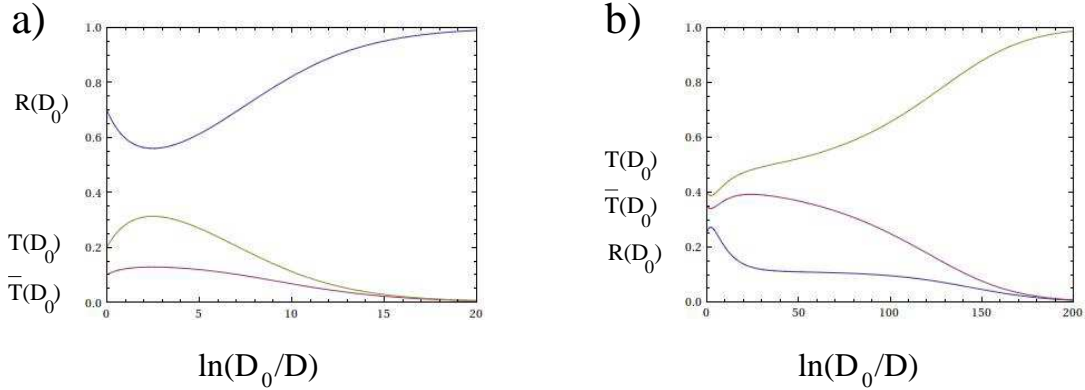}
\caption{Renormalization group flow of the scattering coefficients for a three-wire
junction for non-constant $\beta$:  \\
{\bf (a)}   Renormalization group flow corresponding to  
$T ( D_0 ) = 0.2, \bar{T} ( D_0 ) = 0.1 , R ( D_0 ) = 0.4$ and to
$g_{ 1 , \parallel} (D_0 )  / (2 \pi v ) = 
g_{ 1 , \perp}(D_0 )  / (2 \pi v ) = g_{ 2 , \parallel}  (D_0 )/ (2 \pi v )
= g_{ 1 , \perp }(D_0 ) / (2 \pi v ) = 0.4$. For these values of 
the bare parameters the junction is attracted by the disconnected fixed point
($R \to 1$ while   $T , \bar{T}  \to 0$); \\
 {\bf (b)}   Renormalization group flow corresponding to  
 $T ( D_0 ) =0.4 ,   \bar{T} ( D_0 ) = 0.35 , R ( D_0 ) = 0.25$ and 
 to $g_{ 1 , \parallel} (D_0 )  / (2 \pi v ) = 0.3 ,  
g_{ 1 , \perp}(D_0 )  / (2 \pi v ) = - 0.2 , g_{ 2 , \parallel}  (D_0 )/ (2 \pi v )
= g_{ 1 , \perp }(D_0 ) / (2 \pi v ) = 0.2$.
For these values of 
the bare parameters the junction is attracted by 
the $\chi_{++}$-fixed point.} \label{plot_3j_3}
\end{figure}

For a weak boundary interaction, assuming total spin conservation at the junction, the most relevant boundary interaction
Hamiltonian is a combination of the bilinear operators in  Eq. (\ref{appe.bou1}). The 
relevant scattering processes at the junction are all encoded in the single-particle 
$S$-matrix elements, $S_{ ( j , \sigma) , ( j' , \sigma' ) } ( k )$. Because of spin conservation,
the $S$-matrix is  diagonal 
in the spin index, that is, $S_{ ( j , \sigma ) , ( j' , \sigma' ) } ( k ) = 
\delta_{\sigma , \sigma'} S_{  j ,   j'   } ( k ) $. Assuming also that the boundary Hamiltonian is symmetric 
under exchanging the wires with each other, 
the $3 \times 3$ matrix $S ( k )$ takes the form 
\beq
S  =\left[ \begin{array}{ccc}
r & \bar{t} & t\\
t & r & \bar{t}\\
\bar{t} & t & r
\end{array}\right]
\:\:\:\:.
\label{j3w.3}
\eneq
\noindent
(Note that, in Eq. (\ref{j3w.3}) we assume that all the amplitudes are computed at 
the Fermi level and accordingly drop the index $k$ from the $S$-matrix elements. 
This is consistent with the discussion of Appendix \ref{frg_derivation}, where 
we assume that, close to the Fermi level,  the scattering amplitudes  are
smooth functions of $k$.) It is worth mentioning that, 
in writing Eq. (\ref{j3w.3}), we allowed for time-reversal
symmetry breaking as a consequence, for instance, of a magnetic flux $\phi$ piercing 
the centre of the junction (see Fig. \ref{3wj}). This implies that, in general, 
the scattering amplitude $t$ from wire $j$ to 
wire $j+1$ is different from the one from wire $j$ to wire 
$j-1$ ($\bar{t}$). From Eq. (\ref{j3w.3}) one therefore finds that 
the $F$-matrix elements in Eq. (\ref{fer.15}) are given by 
 $F_{ j , j'} = \frac{\beta}{2} r \delta_{ j , j'}$, with  
again $\beta=\frac{1}{2\pi v}
\left[-g_{1\parallel}-g_{1\perp}+g_{2\parallel}\right]$. 
On applying the FRG formalism of Appendix \ref{frg_derivation}, one readily
obtains the RG-equations for the independent $S$-matrix elements, given by
 
\begin{eqnarray}
\frac{dr}{d\ell} & = & 
\frac{\beta}{2}\left[r-\left|r\right|^{2}r-2t\bar{t}r^{*}\right]
\nonumber \\
\frac{dt}{d\ell} & = & -
\frac{\beta}{2}\left[2t\left|r\right|^{2}+\bar{t}^{2}r^{*}\right]
\nonumber \\
\frac{d\bar{t}}{d\ell} & = & 
-\frac{\beta}{2}\left[2\bar{t}\left|r\right|^{2}+t^{2}r^{*}\right]
\:\:\:\: ,
\label{j3w.4}
\end{eqnarray}
\noindent
which, again, must
be supplemented with the RG-equations for the running coupling 
strengths, Eqs. (\ref{eq:fermionic4},\ref{a.7}). Since, as a consequence of spin conservation in 
scattering processes at the junction, the $S$-matrix is diagonal in the spin indices, 
the charge- and spin-conductance tensors are equal to each other at the fixed points, as well 
as along the RG-trajectories obtained integrating Eqs. (\ref{j3w.4}). In particular, 
using the formalism of Appendix \ref{junoninter}, one obtains 

\beq
G_{ c , s  } ( D ) =  \frac{e^2}{ \pi}  \left[ \begin{array}{ccc}
- R ( D ) + 1 & - \bar{T} ( D ) & -T ( D ) \\  - T ( D )   &   -  R ( D ) + 1 &  - \bar{T} ( D ) 
\\  - \bar{T} ( D ) & -  T ( D )   &     -  R ( D ) + 1 
                            \end{array} \right]
\;\;\;\;,
\label{cflow}
\eneq
\noindent
with $R ( D ) = | r ( D ) |^2 , T ( D ) = | t ( D ) |^2 $, and $\bar{T} ( D ) = | \bar{t} ( D ) |^2$. 
The RG-flow of the scattering coefficients $ T ( D )  , \bar{T} ( D )$ is recovered by solving the 
set of differential equations 
\begin{eqnarray}
\frac{d T}{d \ell} & = & -
\frac{\beta}{2}\left[\left(5 T -\bar{T} \right) ( 1 - T - \bar{T}) -
T \bar{T} \right]\nonumber \\
\frac{d\bar{T} }{d \ell} & = & -
\frac{\beta}{2}\left[\left(5\bar{T}  - T\right)( 1 - T - \bar{T}) -T \bar{T} \right]
\;\;\;\; ,
\label{j3w.5}
\end{eqnarray}
\noindent
which are derived from Eqs. (\ref{j3w.4}) by taking into account the unitarity 
constraint $T (D)  + \bar{T} (D)  + R (D)  = 1$. 
The fixed points of the boundary phase diagram are, therefore, determined by
setting to zero the terms at the right-hand side of Eqs. (\ref{j3w.5}). From 
Eq. (\ref{cflow}) one may therefore recover the corresponding   charge- and the spin-conductance 
tensors.   Borrowing the 
labels used in Ref. [\onlinecite{chamon}], we obtain the following 
fixed points:
 
\begin{itemize}
 \item {\it The $[N_c , N_s]$ (''disconnected'') fixed point}
 
 This fixed point corresponds to having $R= 1 $ and $T = \bar{T} = 0$  which, according to 
 Eq. (\ref{cflow}),   yields 
 
 \beq
 G_c = G_s=  \frac{e^2}{ \pi}  \left[ \begin{array}{ccc}
                                   0 &0 & 0 \\ 0 & 0 & 0 \\ 0 & 0 & 0 
                                  \end{array} \right] 
\:\:\:\: ,
\label{f.4}
\eneq
\noindent
as it is appropriate for a disconnected junction. 

\item {\it The $\chi_{++}$ fixed point}

This corresponds to $ R = \bar{T} = 0$, $T = 1$, which yields

\beq
| S_{ j , j'} |^2 = \left[ \begin{array}{ccc}
0 & 0 & 1 \\ 1 & 0 & 0 \\ 0 & 1 & 0                             
                           \end{array} \right]
\:\:\:\: ,
\label{f.5}
\eneq
\noindent
and, accordingly 

\beq
 G_c = G_s=   \frac{e^2}{ \pi}  \left[ \begin{array}{ccc}
1 & 0 & - 1 \\ - 1 & 1 & 0 \\ 0 & - 1 & 1                             
                           \end{array} \right]
\:\:\:\: ; 
\label{f.6}
\eneq
\noindent

\item {\it The $\chi_{--}$ fixed point}

This corresponds to $ R = T = 0$, $\bar{T} = 1$, which yields

\beq
| S_{ j , j'} |^2 = \left[ \begin{array}{ccc}
0 & 1 & 0 \\ 0 & 0 & 1 \\ 1 & 0 & 0                             
                           \end{array} \right]
\:\:\:\: ,
\label{f.7}
\eneq
\noindent
and, accordingly 

\beq
 G_c = G_s=      \frac{e^2}{ \pi}  \left[ \begin{array}{ccc}
1 & -1 & 0 \\ 0 & 1 & -1 \\ -1 & 0 & 1                             
                           \end{array} \right]
\:\:\:\: ; 
\label{f.8}
\eneq
\noindent

 \item {\it The $M$ fixed point}

This corresponds to $ T = \bar{T} = \frac{4}{9}$, $R = \frac{1}{9}$ and 
has to be identified with the symmetric \cite{lal_1}, or with the Griffith
\cite{griffith} fixed point of a junction of three interacting wires. One obtains

\beq
| S_{ j , j'} |^2 = \left[ \begin{array}{ccc}
\frac{1}{9} & \frac{4}{9} & \frac{4}{9} \\ \frac{4}{9} & 
\frac{1}{9} & \frac{4}{9} \\ \frac{4}{9} & \frac{4}{9} & \frac{1}{9}                             
                           \end{array} \right]
\:\:\:\: ,
\label{f.9}
\eneq
\noindent
and, accordingly 

\beq
 G_c = G_s=   \frac{e^2}{ \pi}  \left[ \begin{array}{ccc}
\frac{8}{9}    & -\frac{4}{9}  & -\frac{4}{9}  \\ -\frac{4}{9}  & \frac{8}{9}   & -\frac{4}{9} 
\\ -\frac{4}{9}  & -\frac{4}{9}  & \frac{8}{9}                            
                           \end{array} \right]
\:\:\:\: . 
\label{f.10}
\eneq
\noindent
\end{itemize}
$\chi_{\pm\pm}$ must clearly be identified with the ''chiral'' fixed points 
of Ref. [\onlinecite{chamon}], where  time-reversal symmetry breaking is maximum, both 
in the charge and in the spin sector. Along the RG-trajectories 
connecting two fixed points, the conductance flow is determined by Eq. (\ref{cflow}). 
The topology and the direction of the RG-trajectories   depend  
on both $\beta$ and on the bare values of the $S$-matrix elements. 
$\beta$ scales with $\ell$ as determined by Eqs. (\ref{eq:fermionic4},\ref{a.7}), 
which makes it necessary to resort to a full numerical integration approach. 
A set of simplified situations can be realized, however, where $\beta$ keeps constant 
along RG-trajectories. For instance, if $g_{1 , \perp} (D_0) = 0$,    Eq. (\ref{a.7}) 
implies that $\beta$ is constant. In this case, from Eqs. (\ref{j3w.5}) 
one readily sees that, if $\beta > 0$, the boundary flow is towards the $NN$-fixed point. At variance, 
if $\beta < 0$ and $T ( D_0 ) > (< ) \bar{T} ( D_0 )$, the boundary flow 
is towards the $\chi_{++}$ ($\chi_{--}$)-fixed point. As an example of possible  RG-trajectories 
that may be realized in this specific case,   in Fig. \ref{plot_3j_1} we plot $ T ( D )$
and $\bar{T} ( D )$ for $\beta $ constant and negative, while we draw similar plots in 
Fig. \ref{plot_3j_2}  for constant and positive $\beta $ and in Fig. \ref{plot_3j_3} 
for non-constant $\beta $ 
(see the captions for details).
\noindent

\noindent

\noindent
All the analysis we have done so far applies to  a junction of three spinful quantum wires
for weak bulk interaction. We now employ the dual-fermion approach  to generalize the FRG-technique to 
regimes corresponding to  strong bulk interactions either in the charge-, or in 
the spin-channel (or in both of them).

\subsection{Fermionic analysis of the strongly interacting regime at $g_c \sim g_s \sim 3$}
\label{gg33}

A first regime to which our dual-fermion approach can be successfully applied 
corresponds to a strong attractive interaction, both in the charge and in the 
spin channels. In particular, we assume $g_c \sim g_s \sim 3$. According to the 
phase diagram derived in Ref. [\onlinecite{chamon}] within the bosonization 
approach, in this range of values of the Luttinger parameters one expects 
to find  a fixed point where paired electron tunneling  and Andreev
reflection are the dominant scattering processes at the junction and, in addition, 
two fixed points with maximally broken time-reversal symmetry, to be identified with 
the $\chi_{++}$ and the $\chi_{--}$-fixed points discussed in the previous subsection.
To apply the FRG-approach to this part of the phase diagram, we have to define 
the appropriate dual fermion coordinates. To do so, let us set 
$g_c = g_s = 3$. We therefore note that, though, in general, the charge- and 
spin-velocities $u_c$ and $u_s$ can be different from each other, one may
easily make them equal by a pertinent rescaling of the real-space coordinate
in the charge- and in the spin-sector of the bosonic Hamiltonian in Eq. (\ref{an.13}). 
As the rescaling does not affect the boundary interaction (which is localized at 
$x=0$), in the following, without loss of generality, we will assume 
$u_c = u_s \equiv u$. In choosing the appropriate dual fermion coordinates, 
we use the criterion of mapping the fixed point we recover in the strongly
interacting limit one-to-one onto those of the phase diagram in the 
weakly interacting regime. Referring to the bosonization formulas of Appendix 
\ref{boso_analysis}, we define the dual bosonic fields $\tilde{\Phi}_{ c (s) , j } ( x  ) , 
\tilde{\Theta}_{ c (s) , j } ( x )$ in terms of those in Eqs. (\ref{an.2},\ref{an.2k}) as

\beq
\left[ \begin{array}{c}
        \tilde{\Phi}_{ c ( s ) , 1 } ( x ) \\
        \tilde{\Phi}_{ c ( s ) , 2 } ( x ) \\
        \tilde{\Phi}_{ c ( s ) , 3 } ( x ) \\
         \tilde{\Theta}_{ c ( s ) , 1 } ( x ) \\
        \tilde{\Theta}_{ c ( s ) , 2 } ( x ) \\
        \tilde{\Theta}_{ c ( s ) , 3 } ( x )
       \end{array} \right] = 
       \left[ \begin{array}{cccccc}
               \frac{1}{\sqrt{3}} &  \frac{1}{\sqrt{3}} &  \frac{1}{\sqrt{3}} & 
               0 & \frac{1}{3} & - \frac{1}{3} \\
                \frac{1}{\sqrt{3}} &  \frac{1}{\sqrt{3}} &  \frac{1}{\sqrt{3}} & 
             - \frac{1}{3}&   0 & \frac{1}{3}  \\
              \frac{1}{\sqrt{3}} &  \frac{1}{\sqrt{3}} &  \frac{1}{\sqrt{3}} & 
              \frac{1}{3}&   -\frac{1}{3} &0 \\
0 & - 1 & 1 & \frac{1}{3 \sqrt{3}} &  \frac{1}{3 \sqrt{3}} &  \frac{1}{3 \sqrt{3}}  \\
1&0 & -1 & \frac{1}{3 \sqrt{3}} &  \frac{1}{3 \sqrt{3}} &  \frac{1}{3 \sqrt{3}}  \\
-1&1 & 0 & \frac{1}{3 \sqrt{3}} &  \frac{1}{3 \sqrt{3}} &  \frac{1}{3 \sqrt{3}}  \\
              \end{array} \right] \left[ \begin{array}{c}
        \Phi_{ c ( s ) , 1 } ( x ) \\
        \Phi_{ c ( s ) , 2 } ( x ) \\
        \Phi_{ c ( s ) , 3 } ( x ) \\
         \Theta_{ c ( s ) , 1 } ( x ) \\
        \Theta_{ c ( s ) , 2 } ( x ) \\
        \Theta_{ c ( s ) , 3 } ( x )
       \end{array} \right]
 \:\:\:\: . 
 \label{strint.1}
 \eneq
 \noindent
Consistently with Eqs. (\ref{an.3}), we therefore define the 
dual fermionic fields as 

\begin{eqnarray}
 \tilde{\psi}_{R , \sigma , j } ( x ) &=& \eta_{R , \sigma , j } 
 e^{ \frac{i}{2} [ \tilde{\Phi}_{ j , c } ( x ) + 
 \tilde{\Theta}_{ j , c } ( x ) + \sigma (  \tilde{\Phi}_{ j , s } ( x ) + 
  \tilde{\Theta}_{ j , s } ( x ) )]} \nonumber \\
  \tilde{\psi}_{L , \sigma , j } ( x ) &=& \eta_{L , \sigma , j } e^{ \frac{i}{2} [  
  \tilde{\Phi}_{ j , c } ( x ) - 
  \tilde{\Theta}_{ j , c } ( x ) + \sigma (  \tilde{\Phi}_{ j , s } ( x ) - 
 \tilde{\Theta}_{ j , s } ( x ) )]} 
 \;\;\;\; . 
 \label{strint.2}
\end{eqnarray}
\noindent
Using Eq. (\ref{strint.1}) one sees that, when expressed in terms of 
the $\tilde{\Phi}$ and of the $\tilde{\Theta}$-fields, the 
bulk Hamiltonian in Eq. (\ref{an.13}) reduces back to the one with 
$g_c = g_s = 1$, which, when expressed in terms of the fermionic
fields defined in Eqs. (\ref{strint.2}), corresponds to the 
free Hamiltonian $H_{0, F}$, given by

\beq
H_{ 0 , F } = - i u \: \sum_{ j  =1}^3 \: \sum_\sigma \:
\int_0^L \: d x \: \left\{ \tilde{\psi}_{ R , j , \sigma}^\dagger ( x ) 
\partial_x  \tilde{\psi}_{ R , j , \sigma}  ( x )  - 
\tilde{\psi}_{ L , j , \sigma}^\dagger ( x ) 
\partial_x  \tilde{\psi}_{ L , j , \sigma}  ( x ) \right\}
\:\:\:\: . 
\label{strint.3}
\eneq
\noindent
Equation (\ref{strint.3}) is the striking result of our technique of introducing 
dual fermion operators: it is a free-fermion Hamiltonian which describes 
a system that is strongly interacting in the original coordinates. Based upon
the dual fermion fields in Eqs. (\ref{strint.2}) one may therefore introduce 
dual boundary operators analogous to those defined in Eq. (\ref{appe.bou1}), namely,
one may set

\beq
\tilde{B}_{(j , j') , \sigma , (X , X' ) }( 0 ) 
= \tilde{\psi}_{ X , j , \sigma}^\dagger ( 0 ) \tilde{\psi}_{X' , j' , \sigma } ( 0 ) 
\:\:\:\: , 
\label{appe.boustrinta}
\eneq
\noindent
and assume that the boundary interaction is realized as a linear combination of the 
operators in Eq. (\ref{appe.boustrinta}) and/or of products of two of them. 
In the absence of additional bulk interaction involving the dual fermion fields,
or in the weakly interacting regime, the most relevant boundary interaction term
is realized as a linear combination of the $\tilde{B}$-operators only. Therefore, 
the physically relevant processes at the junction are all encoded within the 
single-particle $S$-matrix elements in the basis of the dual fields, $\tilde{S}_{ ( j , \sigma) ;
(j' , \sigma' ) }$. A nontrivial  flow for the $\tilde{S}$-matrix elements   
is induced by a nonzero  bulk interaction in the dual-fermion theory, that is, 
by having  $g_{ c (s)} = 3 + \delta g_{ c (s)}$, with $ | \delta g_{c (s)} | / g_{c (s)} \ll 1$.
The dual interaction Hamiltonian, $\tilde{H}_{\rm int}$ can be readily recovered
using Eqs. (\ref{strint.1},\ref{strint.2}). The result is 

\beq
\tilde{H}_{\rm int} =  \sum_{ j = 1}^3 \: \sum_{ \sigma , \sigma'} \: 
g_{ j ; ( \sigma , \sigma')}  \: \int_0^L \: d x \: 
\tilde{\rho}_{ R , j , \sigma} ( x ) \tilde{\rho}_{ L , j , \sigma'} ( x )  \:
+ \sum_{ j\neq j'  = 1}^3 \: \sum_{ \sigma , \sigma'} \: 
g_{ (j , j')  ; ( \sigma , \sigma')} \: \int_0^L \: d x \: 
\tilde{\rho}_{ R , j , \sigma} ( x ) \tilde{\rho}_{ L , j' , \sigma'} ( x ) 
 \:\:\:\: , 
 \label{strint.4}
 \eneq
 \noindent
 with $ \tilde{\rho}_{ R (L) , j , \sigma} ( x ) = 
 : \tilde{\psi}^\dagger_{ R ( L) , j , \sigma } ( x ) 
 \tilde{\psi}_{ R ( L) , j , \sigma } ( x ) :$, and 
 
 \begin{eqnarray}
  g_{ j ; ( \sigma , \sigma')} &=& \frac{2 \pi u ( \delta g_c + \delta g_s) }{9} 
  \delta_{ \sigma , \sigma'}  + \frac{2 \pi u ( \delta g_c - \delta g_s) }{9}
  \delta_{ \sigma , \bar{\sigma}'} \nonumber \\
   g_{ (j , j')  ; ( \sigma , \sigma')} &=& - \frac{8  \pi u ( \delta g_c + \delta g_s) }{9}
  \delta_{ \sigma , \sigma'} - \frac{8  \pi u ( \delta g_c - \delta g_s) }{9} 
  \delta_{ \sigma , \bar{\sigma}'} 
  \:\:\:\: ,
  \label{strint.5}
 \end{eqnarray}
\noindent
plus terms that do not renormalize the scattering amplitudes. 
$\tilde{H}_{\rm int}$ takes the form of the generalized bulk Hamiltonian in 
Eqs. (\ref{fer.a1},\ref{fer.a2}). Given the corresponding $F$-matrix elements 
reported in Eq. (\ref{fer.a4}), one may derive the RG-equations in the case 
in which the spin is conserved at a scattering process at the junction, which 
implies  $\tilde{S}_{ ( j , \sigma ) ; ( j' , \sigma' )  } = 
 \delta_{ \sigma , \sigma'} \tilde{S}_{ j , j' }$, and the boundary 
 interaction is symmetric under a cyclic permutation of the three wires, that is,
 the $\tilde{S}_{ j , j' }$-matrix elements are given by

 \beq
 \tilde{S} = \left[ \begin{array}{ccc}
\tilde{r} & \tilde{\bar{t}} & \tilde{t} \\
\tilde{t} & \tilde{r} & \tilde{\bar{t}} \\
\tilde{\bar{t}} &\tilde{t} & \tilde{r} 
                  \end{array} \right]
\:\:\:\: . 
\label{rr.rf11}
\eneq
\noindent
Summing over both the inter-wire and the intra-wire processes allowed by the bulk interaction, 
one obtains

\begin{eqnarray} 
 \frac{d \tilde{r}}{d \ell} &=& \left( \frac{\gamma - \alpha}{2} \right) 
 \{ \tilde{r} - | \tilde{r} |^2 
 \tilde{r} - 2 \tilde{t} \tilde{\bar{t}} \tilde{r}^* \} \nonumber \\
  \frac{d \tilde{t}}{d \ell} &=& - \left( \frac{\gamma  - \alpha}{2} \right) 
 \{ 2 \tilde{t} | \tilde{r} |^2 + ( \tilde{\bar{t}} )^2 \tilde{r}^* \}
 \nonumber \\
  \frac{d \tilde{\bar{t}}}{d \ell} &=& - \left( \frac{\gamma - \alpha}{2} \right) 
  \{ 2 \tilde{\bar{t}} | \tilde{r} |^2 + ( \tilde{t} )^2 \tilde{r}^* \}
  \:\:\:\: ,
  \label{rr.rf14}
\end{eqnarray}
\noindent
with $\alpha =   g_{ (j , j')  ; ( \sigma , \sigma)} / ( 2 \pi u ) , 
\gamma =   g_{ j   ; ( \sigma , \sigma)} / ( 2 \pi u ) $.  
Equations (\ref{rr.rf14}) are equal with Eqs. (\ref{j3w.4}) in the 
weakly interacting case, provided one substitutes the (running) 
parameter $\beta$ in Eqs. (\ref{j3w.4})  with the (constant)
parameter $\gamma - \alpha$. As a result, the  
RG-flow of the $\tilde{S}$-matrix elements is the same as 
the one obtained for the $S$-matrix elements. Nevertheless, 
due to the nonlinear correspondence between the original and the 
dual fermionic fields, the result for the conductance at corresponding 
points of the phase diagram is completely different. To discuss
this point, let us write the  current operators at fixed spin 
polarization,  $J_{ j , \sigma} ( x )$,  in terms of the dual fermionic fields as

\beq
J_{ j , \sigma} ( x ) = e u  \sum_{ X = L , R } \: \{ 
  \tilde{\rho}_{ X , j-1 , \sigma}  ( x )  -    \tilde{\rho}_{ X , j+1 , \sigma}( x ) 
  \}
  \:\:\:\: ,
  \label{strint.4h}
  \eneq
  \noindent
 with $j + 3 \equiv j$,  and use the formalism of Appendix \ref{appe33} to derive the 
  conductance tensor. From Eq. (\ref{33a.4}) one eventually obtains 
  
  \beq
  G_c ( D ) = G_s ( D ) = G_0   - \frac{3 e^2}{2 \pi} \: \Gamma  (D  ) 
  \;\;\;\; , 
  \label{strint.5h}
  \eneq
  \noindent
  with 
  
  \beq
  G_0 = \frac{e^2}{ \pi} \: \left[ \begin{array}{ccc}
                                     4 & - 2 & - 2 \\ - 2 & 4 & - 2 \\ 
                                     - 2 & - 2 & 4 
                                    \end{array} \right]
\;\;\; , \;\;
\Gamma ( D ) = \left[ \begin{array}{ccc}
                ( \tilde{T} ( D ) + \tilde{\bar{T}} ( D ) ) & -  \tilde{\bar{T}} ( D )
                & - \tilde{T} ( D ) \\ - \tilde{T} ( D ) &
                ( \tilde{T} ( D ) + \tilde{\bar{T}} ( D ) ) & -  \tilde{\bar{T}} ( D ) \\
  -  \tilde{\bar{T}} ( D )  &     - \tilde{T} ( D ) &   
  ( \tilde{T} ( D ) + \tilde{\bar{T}} ( D ) )      
               \end{array} \right]
\:\:\:\: , 
\label{strint.6h}
\eneq
\noindent
and $\tilde{T} (D ) = | \tilde{t} ( D ) |^2 , \tilde{\bar{T}} ( D ) = 
| \tilde{\bar{t}} ( D ) |^2$. The RG-flow of $\tilde{T} (D )$
and of $\tilde{\bar{T}} ( D )$ is determined by Eqs. (\ref{j3w.5}), 
with $\beta$ replaced by $\gamma - \alpha \propto \delta g_c + \delta g_s$. 
As a result, for $\delta g_c + \delta g_s > 0$ the stable fixed point
is the ''dual'' disconnected fixed point, which we dub $[ \tilde{N}_c , 
\tilde{N}_s ]$, as, in bosonic coordinates, it corresponds to imposing Neumann
boundary conditions on all the $\tilde{\Phi}_{ j , c(s) }(x)$-fields at 
$x = 0$. From Eqs. (\ref{strint.5h},\ref{strint.6h}) one therefore finds 
that the corresponding charge- and spin-conductance tensors are given by
$G_c = G_s = G_0$. This is absolutely consistent with the  result provided in 
Ref. [\onlinecite{chamon}] for $g_c = g_s = 3$. Indeed, from Eqs. (\ref{strint.1},\ref{strint.2})
one sees that, resorting back to the original bosonic fields, the $[ \tilde{N}_c , 
\tilde{N}_s ]$-fixed points corresponds to the $[ D_c , D_s ]$-fixed point of 
Ref. [\onlinecite{chamon}], with Dirichlet boundary conditions imposed on 
the relative fields $\varphi_{ 1 , c(s)} ( x ) = \frac{1}{\sqrt{2}} [ \Phi_{ 1 , c (s)} ( x ) 
-  \Phi_{ 2 , c (s)} ( x ) ]$ and $\varphi_{ 2 , c(s)} ( x ) = \frac{1}{\sqrt{6}} 
[ \Phi_{ 1 , c (s)} ( x ) +  \Phi_{ 2 , c (s)} ( x ) - 2  \Phi_{ 3 , c (s)} ( x )  ]$,
where the charge- and the spin-conductance tensors for $g_c = g_s = 3$ 
are equal to each other and both equal to $G_0$. To double-check the consistency
between our dual-fermion FRG-formalism and the bosonization approach, we
note that, at the $ [ \tilde{N}_c , \tilde{N}_s ]$-fixed point, a generic 
linear combination of the boundary operators in Eq. (\ref{appe.boustrinta}) can 
be expressed, in the original bosonic degrees of freedom, as a linear combination of the 
operators  $O_{j , \sigma}  ( 0 )  = 
\eta_{ L , 2 , \sigma } \eta_{ L , 1 , \sigma } \: e^{ - \frac{i}{2}
[ \Theta_{ j+1 , c } ( 0 ) + \Theta_{j , c } ( 0 ) ]  - \frac{i \sigma}{2}
[ \Theta_{ j+1 , s } ( 0 ) + \Theta_{j , s } ( 0 ) ] }$ 
and of their Hermitean conjugates, which is the result obtained 
in Ref. [\onlinecite{chamon}] by means of a pertinent application of the 
delayed evaluation of boundary conditions (DEBC)-technique \cite{oca1,oca2}.

When $\delta g_c + \delta g_s < 0$, the $ [ \tilde{N}_c ,\tilde{N}_s]$-fixed point
becomes unstable. As in the weakly interacting case, we see that, if 
$\tilde{T} ( D_0 ) \neq \tilde{\bar{T}} ( D_0 )$, the junction flows towards
either one of the ''dual-chiral'' fixed points, $\tilde{\chi}_{++} , \tilde{\chi}_{--}$,
with the crossover of the conductance tensors with the scale being given by
Eq. (\ref{strint.5h}). In particular, if $\tilde{T} ( D_0 ) > \tilde{\bar{T}} ( D_0 )$,
the flow is towards the infrared stable $\tilde{\chi}_{++}$-fixed point.
This corresponds to  $\tilde{R} = \tilde{\bar{T}} = 0 , \tilde{T} = 1$. 
From Eq. (\ref{strint.5h}), one therefore obtains that the fixed point
conductances are given by

\beq
G_c = G_s = \frac{e^2}{\pi} \left[ \begin{array}{ccc}
1 & - 2 & 1 \\ 1 & 1 & - 2 \\ - 2 & 1 & 1                     
                   \end{array} \right] \equiv  \frac{e^2}{\pi}
                   {\bf Q}_\chi^+
\;\;\;\; . 
\label{33cond.2}
\eneq
\noindent
By consistency, one would expect that the $\tilde{\chi}_{++}$ fixed point 
should be identified with the $\chi_{++}$ fixed point emerging from the 
weak interaction calculation of the previous section. However, in 
order to compare the conductances obtained in  Eq. (\ref{33cond.2}) with those 
of Eq. (\ref{f.6}) one has to take into account that formula for the conductance tensor
derived within dual-fermion approach applies to a junction connected to 
reservoirs with $g_c = g_s = 3$. Therefore, to make the comparison, one has 
to trade Eq. (\ref{33cond.2}) for a formula for the conductance tensors of a junction 
connected to reservoirs with $g_c = g_s = 1$,  $G_{c ; wl} ,  G_{s ; wl}$. 
As discussed in Appendix \ref{appe33}, this can be done by using Eq. (\ref{ltr.13})
with  $G_{{\rm in} , c(s) } = \frac{2 g_{c(s) }}{g_{c(s)} - 1}
\frac{e^2}{ \pi}$. The result is

\beq
G_{c ; wl  } = G_{s ; wl  } = \frac{e^2}{\pi} \: {\bf Q}_\chi^+ \left\{ {\bf I } 
+ \frac{ {\bf Q}_\chi^+ }{3} \right\}^{-1} = 
 \frac{e^2}{\pi}  \left[ \begin{array}{ccc}
1 & 0 & - 1 \\ - 1 & 1 & 0 \\ 0 & - 1 & 1                             
                           \end{array} \right]
\:\:\:\: , 
\label{22cond.3}
\eneq
\noindent
that is, the same result as in Eq. (\ref{f.6}). Similarly, one can prove that 
the chiral $\tilde{\chi}_{--}$-fixed point, towards which the RG-trajectories flow
if $\tilde{T} ( D_0 ) < \tilde{\bar{T}} (D_0 )$, has to be identified with 
the $\chi_{--}$-fixed point of Sec. \ref{wir}.

When $\tilde{T} (D_0 ) = \tilde{\bar{T}} ( D_0 )$, the RG-trajectories flow
towards a nontrivial fixed point, which we dub $\tilde{M}$, by analogy to the 
$M$-fixed point we found in  Sec. \ref{wir}. Such a fixed point 
 corresponds to $\tilde{R} = 1/9 , \tilde{T} = \tilde{\bar{T}} = 
4 /9$. Therefore, from Eq. (\ref{strint.5h}), one finds that the 
fixed point conductance tensors are given by 

\beq
G_c = G_s = \frac{e^2}{\pi} \: \left[ 
\begin{array}{ccc} 
\frac{4}{3} & - \frac{2}{3} &  - \frac{2}{3} \\
 - \frac{2}{3} &  \frac{4}{3} & - \frac{2}{3} \\- \frac{2}{3} &  - \frac{2}{3} &  \frac{4}{3}
 \end{array} \right]\equiv  \frac{e^2}{\pi}
                   {\bf Q}_\chi^M
\:\:\:\: . 
\label{33cond.4}
\eneq
\noindent
Performing the same transformation as in Eq. (\ref{22cond.3}), 
one eventually finds  

\beq
G_{c ; wl  } = G_{s ; wl } = \frac{e^2}{\pi} \: {\bf Q}_\chi^M \left\{ {\bf I } 
+ \frac{ {\bf Q}_\chi^M }{3} \right\}^{-1} = 
\frac{e^2}{\pi}  \left[ \begin{array}{ccc}
                         \frac{4}{5} & - \frac{2}{5} & - \frac{2}{5} \\
                        - \frac{2}{5} &  \frac{4}{5} & - \frac{2}{5}  \\
 - \frac{2}{5} & - \frac{2}{5}    &  \frac{4}{5}                    
                        \end{array} \right]
\:\:\:\:. 
\label{33cond.5}
\eneq
\noindent
On comparing Eq. (\ref{33cond.5}) with Eq. (\ref{f.10}) we now see that, 
at odds with what happens with the chiral fixed points, 
the $M$ and the $\tilde{M}$-fixed points cannot be identified with 
each other.  
While we are still lacking a clear explanation for this different behavior
at different fixed points, we suspect that this shows that, while the 
conductance at $\chi_{++}$ as well as the $\chi_{--}$-fixed points are in a sense universal, 
that is, independent of the Luttinger parameters (provided one pertinently 
takes into account the corrections due to different Luttinger parameters
for the reservoirs), the conductance at the $M$-fixed point does 
depend explicitly on the Luttinger parameters. This would definitely not be surprising, 
as such a feature would be
shared by a similar fixed point such as, for instance, the nontrivial fixed 
point at a junction between a topological superconductor and two interacting
one-dimensional electronic systems \cite{GiuAf}.  In any case, we believe 
that this issue calls for a deeper investigation, which will possibly
be the subject of a forthcoming work. 

As  so far we mainly concentrated around  the ''diagonal''  in 
Luttinger parameter plane, that is, at $g_c \sim g_s$, we are now going to complement
our analysis by discussing the regime with $g_c \sim 3 , g_s \sim 1$, together 
with the complementary one, $g_c \sim 1 , g_s \sim 3$.

\subsection{Fermionic analysis of the strongly interacting regime for 
$g_c \sim 3 , g_s \sim 1$ and $g_c \sim 1 , g_s \sim 3$}
\label{gg31}

We now discuss the  ''asymmetric'' regime  $g_c \sim 3 , g_s \sim 1$. 
In order to recover the whole procedure, we again note that it is always 
possible to separately rescale the real-space coordinate in the 
bosonic Hamiltonian in Eq. (\ref{an.13}), so to make the charge- and the spin-plasmon
velocities   to be both equal to $u$. Therefore, to actually define 
the dual fermion coordinates, let us assume $g_c = 3 , g_s = 1$. Due to the absence 
of bulk interaction in the spin sector, we have no need to transform the $\Phi_{ s , j} , 
\Theta_{ s , j}$-fields. At variance, we do trade the fields $\Phi_{ c , j } , 
\Theta_{ c , j }$ for the fields $\tilde{\Phi}_{ c , j } , 
\tilde{\Theta}_{ c , j }$ defined in Eq. (\ref{strint.1}). Accordingly, we  
consistently define   the  dual fermionic fields as 

\begin{eqnarray}
 \chi_{R , \sigma , j } ( x ) &=& \eta_{R , \sigma , j } 
 e^{ \frac{i}{2} [ \tilde{\Phi}_{ j , c } ( x ) + 
 \tilde{\Theta}_{ j , c } ( x ) + \sigma (  \Phi_{ j , s } ( x ) + 
  \Theta_{ j , s } ( x ) )]} \nonumber \\
  \chi_{L , \sigma , j } ( x ) &=& \eta_{L , \sigma , j } e^{ \frac{i}{2} [  
  \tilde{\Phi}_{ j , c } ( x ) - 
  \tilde{\Theta}_{ j , c } ( x ) + \sigma (  \Phi_{ j , s } ( x ) - 
 \Theta_{ j , s } ( x ) )]} 
 \;\;\;\; . 
 \label{strint.2b}
\end{eqnarray}
\noindent
Again, one sees   that, when expressed in terms of 
the dual fermion operators in Eqs. (\ref{strint.2b}),   the 
bulk Hamiltonian in Eq. (\ref{an.13}) reduces back to the free fermionic 
Hamiltonian, given by

\beq
H_{ 0 , F ; \chi  } = - i u \: \sum_{ j  =1}^3 \: \sum_\sigma \:
\int_0^L \: d x \: \left\{ \chi_{ R , j , \sigma}^\dagger ( x ) 
\partial_x  \chi_{ R , j , \sigma}  ( x )  - 
\chi_{ L , j , \sigma}^\dagger ( x ) 
\partial_x  \chi_{ L , j , \sigma}  ( x ) \right\}
\:\:\:\: . 
\label{strint.3b}
\eneq
\noindent
Having defined the dual fermion operators, we now assume that the 
leading boundary perturbation is realized as a linear combination of 
the dual boundary operators defined as

\beq
\tilde{B}_{\chi ; (j , j') , \sigma , (X , X' ) }( 0 ) 
= \chi_{ X , j , \sigma}^\dagger ( 0 ) \chi_{X' , j' , \sigma } ( 0 ) 
\:\:\:\: , 
\label{appe.boustrint}
\eneq
\noindent
 and/or of products of two of them. Just as we have done before, we
also assume that the most relevant  scattering processes 
at the junction are fully described by means of the 
single-particle $S$-matrix elements in the basis of the $\chi$-fields, $S_{ \chi ; 
( j , \sigma) ; (j' , \sigma' ) }$. Slightly displacing $(g_c , g_s)$ from (3,1), that is, 
setting  $g_c = 3 + \delta g_c , g_s = 1 + \delta g_s$, with $ | \delta g_c | / 3 , | \delta g_s | 
\ll 1$, gives rise to an effective interaction Hamiltonian $H_{\chi ; {\rm int}}$, 
which takes exactly the same form as $\tilde{H}_{\rm int}$ in Eq. (\ref{strint.4}),
and is given by

\beq
H_{\chi ; {\rm int}} =  \sum_{ j = 1}^3 \: \sum_{ \sigma , \sigma'} \: 
g_{\chi ;  j ; ( \sigma , \sigma')} \: \int_0^L \: d x \: 
\rho_{ \chi ;  R , j , \sigma} ( x ) \rho_{ \chi ;  L , j , \sigma'} ( x )  \:
+ \sum_{ j\neq j'  = 1}^3 \: \sum_{ \sigma , \sigma'} \: 
g_{  \chi ; (j , j')  ; ( \sigma , \sigma')} \: \int_0^L \: d x \: 
\rho_{ \chi ; R , j , \sigma} ( x ) \rho_{ \chi ; L , j' , \sigma'} ( x ) 
 \:\:\:\: , 
 \label{strint.b4}
 \eneq
 \noindent
 with $ \rho_{ \chi ;  R (L) , j , \sigma} ( x ) = 
 : \chi^\dagger_{ R ( L) , j , \sigma } ( x ) 
 \chi_{ R ( L) , j , \sigma } ( x ) :$, and 
 
 \begin{eqnarray}
  g_{\chi ;  j ; ( \sigma , \sigma')} &=& \frac{8 \pi u ( \delta g_c - 3  \delta g_s) }{9} 
  \delta_{ \sigma , \sigma'}  + \frac{8 \pi u ( \delta g_c +3  \delta g_s) }{9}
  \delta_{ \sigma , \bar{\sigma}'} \nonumber \\
   g_{ \chi ;  (j , j')  ; ( \sigma , \sigma')} &=& - \frac{4  \pi u ( \delta g_c -3 \delta g_s) }{9}
  \delta_{ \sigma , \sigma'} - \frac{4  \pi u ( \delta g_c +3  \delta g_s) }{9} 
  \delta_{ \sigma , \bar{\sigma}'} 
  \:\:\:\: , 
  \label{strint.b5}
 \end{eqnarray}
\noindent
plus terms that do not renormalize the scattering amplitudes. Making 
the assumption that the spin is conserved at a scattering process at 
the junction, we again obtain that $S_{ \chi ; ( j , \sigma ) ; ( j' , \sigma' )  } = 
 \delta_{ \sigma , \sigma'} S_{ \chi ; ( j , j' )  }$, with, for a 
 boundary interaction symmetric under a cyclic permutation of the three wires,
 the  $S_{ \chi  }$-matrix  being given by

 \beq
 S_\chi  = \left[ \begin{array}{ccc}
r_\chi  & \bar{t}_\chi & t_\chi  \\
t_\chi & r_\chi &  \bar{t}_\chi  \\
 \bar{t}_\chi  & t_\chi & r_\chi 
                  \end{array} \right]
\:\:\:\: . 
\label{rr.rf11b}
\eneq
\noindent
The RG-equations for the running $S_\chi$-matrix elements are derived in 
perfect analogy with Eq. (\ref{rr.rf14}). The result is exactly the same, except that 
now $\gamma - \alpha \propto \delta g_c - 3 \delta g_s$.  
In order to trace out the correspondence between the fixed point of 
the boundary phase diagram for the dual-fermion scattering amplitudes and those 
of the phase diagram for the original fermion amplitudes, we 
now discuss the behavior of the charge- and of the spin-conductance tensor 
along the RG-trajectories. To do so, we note that, due to the fact that
the spin sector of the theory is left unchanged, when resorting to the 
dual coordinates, the spin-conductance tensor, when expressed in 
terms of the scattering coefficients at the junction, takes the same
form as in the noninteracting case, given in Eq. (\ref{cflow}). As
variance, the charge-conductance tensor depends on the scattering 
coefficients as given in Eq. (\ref{strint.6h}). As a result, one 
  obtains 

\beq
G_c ( D ) = G_0 - \frac{3 e^2}{2 \pi} \Gamma_\chi ( D ) \;\;\; , \;\;
G_s ( D ) =\frac{ e^2}{\pi}  \left[ \begin{array}{ccc}
                                     - R_\chi ( D ) + 1 & - \bar{T}_\chi ( D ) & 
                                     - T_\chi ( D ) \\ - T_\chi ( D )& 
  - R_\chi ( D ) + 1 & - \bar{T}_\chi ( D ) \\  - \bar{T}_\chi ( D ) &  - T_\chi ( D )& 
   - R_\chi ( D ) + 1  
                                    \end{array} \right]
                                    \;\;\;\; ,
                                    \label{strint.h11}
                                    \eneq
                                    \noindent
with $ R_\chi ( D ) = | r_\chi ( D )|^2 , T_\chi (D ) = | t_\chi ( D )|^2 ,
\bar{T}_\chi ( D ) = | \bar{t}_\chi ( D) |^2$. We are, now, in the position 
of mapping out the whole phase diagram of the spinful junction for 
$g_c \sim 3, g_s \sim 1$, including the fixed point manifold, and of tracing 
out the correspondence between the fixed points given in terms of the dual 
fermion amplitudes, and the   described in terms of the original 
fermionic coordinates \cite{chamon}. First of all, we note that, when 
$\delta g_c - 3 \delta g_s > 0$, the system is attracted towards the ''dual disconnected''
fixed point $[N_{\chi , c} , N_{ \chi , s}]$, characterized by 
the scattering coefficients $R_\chi = 1 , T_\chi = \bar{T}_\chi = 0$. At such 
a fixed point, one obtains $G_c = G_0 , G_s = 0$, which enables us to identify 
$[N_{\chi , c} , N_{ \chi , s}]$ with the $[D_c , N_s]$-fixed point in 
the phase diagram of Ref. [\onlinecite{chamon}], that is, with a spin-insulating 
fixed point where the most relevant process at the junction is pair-correlated 
Andreev reflection in each wire. At variance, when $\delta g_c - 3 \delta g_s < 0$,
the junction flows towards one among the dual $\chi_{++}, \chi_{--}$, or $M$-fixed points. 
In particular, from Eqs. (\ref{strint.h11}), one sees that, 
at the dual $\chi_{++}$-fixed point, $G_c$ is given by 
Eq. (\ref{33cond.2}), while $G_s$ takes the form provided in Eq. (\ref{f.6}).
After the correction of Eq. (\ref{22cond.3}), one eventually finds that 
$G_c$ and $G_s$ are equal to each other, and both equal to the fixed-point 
conductance at the $\chi_{++}$-fixed point in the original coordinates. Thus, 
we are eventually led to identify the dual $\chi_{++}$-fixed point with the 
analogous one, realized in the original coordinates. A similar argument 
leads to the identification of the    dual $\chi_{--}$-fixed point with the 
analogous one, realized in the original coordinates. As for what concerns the 
dual $M$-fixed point, after correcting the charge-conductance tensor as 
in Eq. (\ref{33cond.5}), one finds that, at such a fixed point, 

\beq
G_{c ; wl  } =   
\frac{e^2}{\pi}  \left[ \begin{array}{ccc}
                         \frac{4}{5} & - \frac{2}{5} & - \frac{2}{5} \\
                        - \frac{2}{5} &  \frac{4}{5} & - \frac{2}{5}  \\
 - \frac{2}{5} & - \frac{2}{5}    &  \frac{4}{5}                    
                        \end{array} \right] \;\;\; , \;\;
                        G_{s ; wl} = G_s =  \frac{e^2}{ \pi}  \left[ \begin{array}{ccc}
\frac{8}{9}    & -\frac{4}{9}  & -\frac{4}{9}  \\ -\frac{4}{9}  & \frac{8}{9}   & -\frac{4}{9} 
\\ -\frac{4}{9}  & -\frac{4}{9}  & \frac{8}{9}                            
                           \end{array} \right]
\:\:\:\:. 
\label{strint.h12}
\eneq
\noindent 
Putting together Eqs. (\ref{strint.h12},\ref{33cond.5},\ref{f.10}) we 
again see that $M$-like fixed points are not mapped onto each other, 
not even after the correction of Eq. (\ref{33cond.5}). This is again 
consistent with our previous hypothesis, namely, that  the conductance at the $M$-fixed point does 
depend explicitly on the Luttinger parameters and, therefore, it 
is different in the various cases we discussed before.
 
Before concluding this subsection, we  point out that the same analysis 
we just performed in the case $g_c \sim 3 , g_s \sim 1$ does apply equally well to 
the complementary situation $g_c \sim 1 , g_s \sim 3$, provided one  swaps 
charge- and spin-operators (and conductances) with each other. 

Putting together all the results we obtained using the FRG-approach, one may infer 
the global topology of the phase diagram of a spinful three-wire junction and compare
the results with those obtained within the bosonization approach. This will be 
the subject of the next subsection.

\subsection{Global topology of the phase diagram from fermionic renormalization group
approach}
\label{g_phase}

Following Ref. [\onlinecite{chamon}], we discuss the main features of the phase
diagram of the three-wire junction within various regions in the $g_c - g_s$ plane.
Let us start from the ''quasisymmetric'' region $g_c \sim g_s$. From the results of 
Sec. \ref{wir} we see that, setting $g_c = 1 + \delta g_c , 
g_s = 1 + \delta g_s$, as long as $\delta g_c + \delta g_s < 0$ (corresponding to 
$\frac{1}{2g_c} + \frac{1 }{2 g_s} < 1$), the system flows towards the disconnected
$[ N_c , N_s]$-fixed point. At variance, for $\delta g_c + \delta g_s > 0$
(that is, for $\frac{1}{2g_c} + \frac{1 }{2 g_s} > 1$), as soon as 
scattering processes from wire $j$ to wires $j\pm 1$ take place at 
different rates (i.e., $T \neq \bar{T}$), the RG-trajectories flow towards either one
of the $\chi_{++}$ or $\chi_{--}$-fixed points. While this is basically consistent with 
the region of  the  phase diagram derived in  \cite{chamon} corresponding to 
$g_c \sim g_s \sim 1$, in addition, when $T = \bar{T}$, we found that 
the system flows towards a symmetric fixed point, which we dubbed $M$, 
  with peculiar, $g_c , g_s$-dependent
transport properties. Within the FRG-approach we were able to map out the full crossover 
of the charge- and spin-conductance tensor between any two of the fixed points listed above,
with some paradigmatic examples shown in the figures of Sec. \ref{wir}. 
Keeping within the quasisymmetric region, in Sec. \ref{gg33} we show
that, setting $g_c = 3 + \delta g_c , 
g_s = 3 + \delta g_s$, for $\delta g_c + \delta g_s > 0$ (that is, for 
$g_c + g_s > 6$), the stable RG-fixed point corresponds to the  $[ D_c , D_s ]$-fixed point of 
Ref. [\onlinecite{chamon}] while, as soon as $\delta g_c + \delta g_s < 0$ (that is, for 
$g_c + g_s < 6$), the system flows towards either one of the $\chi_{++}$ or $\chi_{--}$
fixed points in the non-symmetric case, or towards an ''$M$-like'' fixed point in 
the symmetric case. As discussed above, while, at both the $\chi_{++}$ and 
the $\chi_{--}$-fixed points the conductance tensors for the junction not connected to 
the leads are the same, regardless of the value of the Luttinger parameters, at variance, at
the $M$-fixed point they do depend on $g_c $ and $g_s$ and, in this sense, they appear to 
be ''nonuniversal''. Over all, the results we obtained across the 
region $g_c \sim g_s$ are consistent with a phase diagram where 
the $[N_c , N_s ]$ and the $[D_c , D_s ]$-fixed points are respectively 
stable for $\frac{1}{2 g_c} + \frac{1}{2 g_s} < 1$ and for 
$g_c + g_s > 6$, while, at intermediate values of the Luttinger parameters, depending 
on the bare values of the scattering coefficients at the junction, one out of the 
(universal) chiral $\chi_{++} , \chi_{--}$-fixed points or the (nonuniversal)
$M$-fixed point becomes stable. This results already complements the phase 
diagram of Ref. [\onlinecite{chamon}]  by introducing the $M$-fixed point, which 
has necessarily to be there, in order to separate the phases corresponding to 
$\chi_{++}$ and to $\chi_{--}$ from each other. To push our analysis outside of 
the $g_c \sim g_s$-region, we discussed the nonsymmetric case $g_c \sim 3 , g_s \sim 1$. 
In this case, we found that the manifold of fixed points consists of the 
$[D_c , N_s ]$ asymmetric fixed point, at which the junction is characterized by 
perfect pair-correlated Andreev reflection in each wire, while it is perfectly 
insulating in the spin sector, the chiral $\chi_{++} , \chi_{--}$-fixed points and, again, 
an $M$-like fixed point. Also in this region our results appear on one hand to be consistent 
with the phase diagram of Ref. [\onlinecite{chamon}], on the other hand to complement it 
with singling out the $M$-like fixed point. The complementary regime 
$g_c \sim 1 , g_s \sim 3$ can be straightforwardly recovered from the previous
discussion by just swapping charge and spin with each other.
Aside from recovering the global phase diagram 
of the junction, our technique allows  for generalizing 
to strongly-interacting regimes the main advantage of using fermionic, rather than bosonic coordinates,
that is, the possibility mapping out the crossover of the 
conductance between fixed points in the phase diagram.

\section{Discussion and conclusions}
\label{conclusions}

In the paper, we  generalize the RG approach to junctions of strongly-interacting QWs.
  In order to do so, we make a combined use of 
both the fermionic and the bosonic approaches to interacting electronic 
systems in one dimension, which enables us to build pertinent nonlocal 
transformation between the original fermion fields and dual-fermion operators,
so that, a theory that is strongly interacting in terms of the former ones, 
maps onto a weakly interacting one,  in terms of the latter ones.
On combining the dual-fermion approach with the FRG-technique, 
we are able to produce new and interesting results,
already  for the well-known two wire junction. When applied to a $Z_{3}$-symmetric 
three-wire junction, our technique first of all 
allows for recovering fundamental informations concerning the topology 
of the global phase diagram, as well as all  the fixed points accessible to the 
junction in various regions of the parameter space. While, in this respect, our approach 
looks like a useful means to complement the bosonization approach to conductance properties 
of junctions of quantum wires, where it appears extremely useful and, in a sense, rather 
unique, is in providing the full crossover of the conductance tensors between any two 
fixed points connected by an RG-trajectory.  The crossover in the conductance properties 
can be experimentally mapped out by monitoring the transport properties of the
junction as a function of a running reference scale, such as the temperature, or the 
effective system size. While the standard FRG approach just yields crossover curves at
weak bulk interaction in the quantum wires \cite{matgla1,matgla2,lal_1,nazarov_glazman}, as stated above, 
our approach extends such a virtue of the fermionic approach to regions at 
strong values of the bulk electronic interaction in the wires. 
 
By resorting to the appropriate  dual fermionic degrees of freedom,
our approach allows for describing in terms of 
effectively one-particle $S$-matrix elements correlated pair scattering 
and/or Andreev reflection, in regions of values of the interaction parameters 
where they correspond to the most relevant scattering processes at the 
junction. This allows for envisaging, within our technique, fixed points such as 
the $[ D_c , D_s]$, or the $[D_c , N_s]$ one. In fact, due to basic assumption of 
the standard FRG-approach that all the relevant processes at the junction are encoded 
in the single-particle $S$-matrix elements, fixed points such as those 
listed before are typically  not expected to be recovered without resorting 
to the appropriate dual fermion coordinates,  not even after relaxing the weak bulk interaction
constraint \cite{aristov_2}. 
 
While, for simplicity, here we restrict ourselves to the case of a symmetric junction and of
a  spin-conserving boundary interaction at  the junction, our approach can be readily 
generalized to a non-symmetric junction characterized, for instance, by different Luttinger 
parameters in different wires \cite{chamon_2}, and/or by a non-spin-conserving 
boundary interaction. Also, a generalization of our approach to a junction involving 
ordinary \cite{belzig}, or topological superconductors \cite{fidkowski,GiuAf} is likely 
to allow for describing the full crossover of the conductance in a single
junction, as well as of the equilibrium (Josephson) current in a SNS-junction thus
generalizing, in this latter case, the results obtained Refs. [\onlinecite{GiuAf2,GiuAf2a,GiuAf3}]
to an SNS-junction with an interacting central region. 

Finally, it is worth stressing that our technique generically complements the RG-approach, 
so to extend it to strongly-interacting 
problems. Very likely, rather than combining it with the FRG-technique, one could work out 
the RG-trajectories by using the alternative (and, to some extent, more accurate, though less
tractable analytically) fRG-approach \cite{meden1,meden2,meden3,meden4,meden5,meden5.1,meden6}. Although 
we believe this is a potentially interesting research topic to pursue, it goes beyond the 
scope of this work, where, for the sake of simplicity and analytical tractability, we rather 
preferred to use the FRG-technique.

\vspace{0.5cm}

We would like to thank I. Affleck and Z. Shi for enlightening discussions and  for 
sharing with us their unpublished results 
about the FRG-approach to a junction of three spinless quantum wires.  
We acknowledge insightful discussions with A. Zazunov and E. Eriksson  at 
various stages of completion of this work. A. N. acknowledges financial 
support from European Commission, European Social Fund
and Regione Calabria.

\appendix

\section{Derivation of the fermionic renormalization group equations 
for the $S$-matrix}
\label{frg_derivation}

In this Appendix we review the  derivation of 
the FRG-equations for the $S$-matrix elements describing 
single-particle scattering at a junction of quantum wires, 
as discussed in Refs. [\onlinecite{matgla1,matgla2,lal_1}]. In our
paper we compute dc-transport properties of junctions of quantum wires.
In doing so, we describe each wire by only retaining its low-energy,  
long-wavelength excitations about the Fermi points $\pm k_F$. Finally, 
we alledge for the system to present an ''inner'' boundary at $x=0$, where
the junction is located and the boundary conditions on the 
fields are determined by the boundary interaction describing the junction, 
and an ''outer'' boundary at $x=L$ which is just required for introducing a
cutoff length scale $L$, eventually sent to $\infty$ at the end of the 
calculations. As a result, the ''bulk'' Hamiltonian for the wires is 
realized as   $H_{\rm Bulk} = H_0 + H_{\rm V}$, 
with $H_0$ being the free Hamiltonian for noninteracting chiral fermions 
in Eq. (\ref{fer.1}), while  the two-body interaction Hamiltonian 
$H_{\rm V}$ is given by 
 
\beq
H_{\rm V} = \sum_{ j = 1}^K 
\frac{1}{2}\sum_{\sigma\sigma'}\int_0^L\: dx \: dy \: 
\rho_{j,\sigma}(x) V_{j ; \sigma , \sigma'}(x-y) \rho_{j,\sigma'}(y)
\:\:\:\: . 
\label{fer.4}
\eneq
\noindent
 Note that, as typically done in this class of problems \cite{matgla1,matgla2,lal_1,nazarov_glazman},
 in Eq. (\ref{fer.4}), we are assuming a purely
''intra-wire'' interaction. In fact, in the analysis of 
our paper, we had to deal with an emerging bulk interaction 
Hamiltonian with nontrivial inter-wire interaction terms. 
Yet, for the sake of simplicity, we prefer to add inter-wire
interactions to the simplified model Hamiltonian, where 
a local approximation for the interaction potential is 
done. Indeed, on assuming a short-range interaction potential effective over a 
typical length scale $\lambda$ (as it happens, for instance, in 
gated wires), one may approximate Eq. (\ref{fer.4}) by setting  
$x\sim y$ in the Coulomb potential, so that $H_{\rm V }$ is 
approximated by $H_{\rm int}$ in Eq. (\ref{eq:fermionic1}).
In order to derive Eq. (\ref{eq:fermionic1}), we have neglected 
terms obtained integrating operators proportional to the rapidly oscillating 
functions $e^{  \pm 2ik_{F}x }$. Based on an analogous argument, 
we assume that the umklapp term ($\propto e^{\pm 4 i k_F x }$)
does not appear either (in fact, an additional argument, related to
the irrelevance of the corresponding operator, can be recovered within 
the bosonization approach to the problem \cite{shultz,gogolin}). 
Finally, it is worth mentioning that we also ignore terms of   the form

\beq
H_{int, g_{4}} =  \frac{1}{2} \; \sum_{ j = 1}^K \: 
\sum_{\sigma\sigma'} \: V_{j,\sigma\sigma'}(0) \:  \int_0^L  \:  
dx \: \left[\rho_{L,j,\sigma} ( x ) 
\rho_{L,j,\sigma'} ( x ) +
\rho_{R,j,\sigma} ( x ) \rho_{R,j,\sigma'} ( x ) \right]
\:\:\:\: , 
\label{fer.8}
\eneq
\noindent
since   they just  renormalize the Fermi velocity and the chemical
potential \cite{shultz} and, so, do not effectively contribute to the renormalization
of the $S$-matrix at the junction. Two relevant   limiting cases  
can be recovered from Eq. (\ref{eq:fermionic1}). The former one 
corresponds to spinless fermions and is recovered by dropping terms 
containing fields with a given spin polarization, say $\sigma = \downarrow$.
In this case, on setting $\psi_{j,R(L),\uparrow }  \to \psi_{j,R(L)  }$, it easy to 
verify that  Eq. (\ref{eq:fermionic1}) reduces to 

\begin{eqnarray}
H_{\rm int} & \to & \sum_{ j = 1}^K \:  \int_0^L  \: dx
\: \left[ g_{j,1}\psi_{R,j}^{\dagger} ( x ) 
\psi_{L , j }^{\dagger} ( x ) \psi_{R , j } (x ) 
\psi_{L , j} ( x ) +
g_{j,2}\psi_{R , j}^{\dagger} ( x ) \psi_{L , j}^{\dagger} ( x ) 
\psi_{L , j} ( x ) \psi_{R , j } ( x ) \right] \nonumber \\
 & = & \sum_{ j = 1}^K 
 (g_{j,2}-g_{j,1}) \;  \int_0^L   \: dx \: \left[\psi_{R , j}^{\dagger} ( x ) 
 \psi_{L , j}^{\dagger} ( x ) \psi_{L , j} ( x ) \psi_{R , j} ( x ) \right]
 \;\;\;\; , 
 \label{fer.9}
\end{eqnarray}
\noindent
in agreement with Eq. (6) of Ref. [\onlinecite{lal_1}]. The latter
one corresponds to assuming spinful fermions, but a   spin
independent interaction, setting $g_{j,1(2)\parallel}=
g_{j,1(2),\perp}=g_{j,1(2)}$. In this case, 
Eq. (\ref{eq:fermionic1}) reduces to

\beq
H_{\rm int} \to 
\sum_{\sigma\sigma'} \:   \int_0^L  \: dx \: \left[
g_{j,1}\psi_{R, j , \sigma}^{\dagger} ( x ) \psi_{L, j , \sigma'}^{\dagger} ( x ) 
\psi_{R, j , \sigma'} ( x ) \psi_{L, j , \sigma} ( x ) +
g_{j,2}\psi_{R, j , \sigma}^{\dagger} ( x ) \psi_{L, j , \sigma'}^{\dagger} ( x ) 
\psi_{L, j , \sigma'} ( x ) \psi_{R, j , \sigma} ( x ) \right]
\;\;\;\; , 
\label{fer.10}
\eneq
\noindent
in agreement with Eq. (50) of  Ref. [\onlinecite{lal_1}]. 
As already stated in the main text, FRG approach typically applies to the case in 
which the boundary scattering processes are fully described by the single-particle
$S$-matrix elements in Eq. (\ref{2wj.1}). 
The first FRG step consists in trading the 
quartic bulk interaction Hamiltonian for a quadratic one, by
means of a pertinent Hartree-Fock (HF) decomposition of the interaction \cite{matgla1,matgla2,lal_1}.
After performing the HF decomposition and  making the standard assumption that, as 
we are dealing with low-energy excitations around the Fermi level, the 
$S$-matrix elements can be taken to be all independent of energy, 
$H_{\rm int}$ reduces to

\begin{eqnarray}
H_{\rm int} & \to  & - \sum_{ j = 1}^K \; 
\frac{i\left(g_{j,2,\parallel}-g_{j,1,\parallel}-g_{j,1,\perp}\right)}{4\pi} \:
 \int_0^L \: \frac{dx}{x}\: \biggl[S_{j,j}^{*}\left(\psi_{L,j , \uparrow}^{\dagger} ( x ) 
\psi_{R, j , \uparrow} (x) + 
\psi_{L, j , \downarrow}^{\dagger} ( x ) \psi_{R, j , \downarrow} ( x ) \right)
\nonumber \\
 & - & S_{j,j}\left(\psi_{R, j ,\uparrow}^{\dagger} (x ) 
 \psi_{L, j , \uparrow} ( x) +
 \psi_{R,j , \downarrow}^{\dagger} ( x ) \psi_{L, j , \downarrow} ( x ) 
 \right)\biggr]
 \:\:\:\: , 
 \label{fer.12}
\end{eqnarray}
\noindent
for spinful electrons, and to

\beq
H_{\rm int}   \to   - \sum_{ j = 1}^K \; 
\frac{i\left(g_{j,2,\parallel}-g_{j,1,\parallel}\right)}{4\pi} \int_0^L \: 
\frac{dx}{x} \: \biggl[S_{j,j}^{*}\psi_{L, j }^{\dagger} ( x ) 
\psi_{R,j} ( x ) -S_{j,j}\psi_{R,j}^{\dagger} ( x ) \psi_{L,j} ( x ) \biggr]
\:\:\:\: , 
\label{fer.13}
\eneq
\noindent
for spinless electrons. The corrections to the amplitude
for an incoming/outgoing electron in wire $j$ to become an outgoing/incoming
electron in wire $j'$ under the effect of $H_{\rm int}$ are readily 
computed using the approach of  \cite{matgla1,matgla2,lal_1}. The corresponding
corrections to the $S_{j,j'}$ matrix elements are given by

\beq
dS_{j,j'}   = - \sum_{ i ,i'=1}^K  \:
\left (S_{j , i}   \left( F^{\dagger}\right)_{i,i'}S_{i' , j'}   -
F_{j , j'}\right) \: d \ell 
\:\:\:\: , 
\label{fer.14}
\eneq
\noindent
with  $d \ell = d\ln(L/\lambda)$, $\lambda$ being a length scale corresponding to the  (finite) range
of the interaction potential   $V_{j,\sigma\sigma'}(x)$ and the Friedel matrix
$F$ defined as the block-diagonal matrix 
 
\beq
F_{j,j'}=\frac{1}{4\pi v}
\left(g_{j,2,\parallel}-
g_{j,1,\parallel}-g_{j,1,\perp}\right)
S_{j,j}\delta_{j,j'} \;\;\;\; , 
\label{fer.15}
\eneq
\noindent
in the spinful case, while 
 
\beq
F_{j,j'}=
\frac{1}{4\pi v}
\left(g_{j,2,\parallel}-g_{j,1,\parallel}\right)
S_{j,j}
\delta_{j,j'}
\:\:\:\: , 
\label{fer.16}
\end{equation}
\noindent
in the spinless case.  The term $\ln(L/ \lambda)$   is due the presence of the $x^{-1}$ 
in the integral giving the correction to the amplitudes within the
HF-approximation. Throughout the poor man's scaling method, one  can replace
 $d \ell = d\ln(L/\lambda)$ with $ d\ln(D_{0}/D)$, with $D_0$ being a high-energy cutoff
 ($\sim$ bandwidth of bulk electrons) and $D$ being the low-energy running 
 scale.  The RG-equations for the $S$-matrix must be supplemented with
the RG-equations for the interaction constants $g_{1 , \parallel}$, $g_{1 , \perp}$,
$g_{2 , \parallel}$, $g_{2 , \perp}$. The  procedure is discussed in 
detail, for  instance, in Ref. [\onlinecite{sugy}]. Here, we just provide the 
final result for the scaling equations of the bulk interaction strengths, which is \cite{sugy} 

\begin{eqnarray}
\frac{dg_{j,2 , \parallel}}{d\ell} & =& -\frac{1}{2\pi v}  (g_{j,1 , \parallel})^{2}\nonumber \\
\frac{dg_{j,2 , \perp}}{d \ell } & =&-\frac{1}{2\pi v}  (g_{j,1 , \perp})^{2}\nonumber \\
\frac{dg_{j,1 , \parallel}}{d\ell} & =&-\frac{1}{2\pi v}   \left[(g_{j,1 , \parallel})^{2}+(g_{j,1 , \perp})^{2}\right]\nonumber \\
\frac{dg_{j,1 , \perp}}{d\ell} & =&-\frac{1}{2\pi v} 
2 g_{j,1 , \perp}\left[g_{j,2 , \perp}-g_{j,2 , \parallel}+g_{j,1 , \parallel}\right]
\:\:\:\: , 
\label{eq:fermionic4}
\end{eqnarray}
\noindent
valid in the weak coupling regime. For a spin-independent interaction,
Eqs. (\ref{eq:fermionic4}) reduce to

\begin{eqnarray}
\frac{dg_{j,2 , \parallel}}{d\ell} &=& -\frac{1}{2\pi v}(g_{j,1 , \parallel})^{2}\\
\frac{dg_{j,1 , \parallel}}{d\ell} &=& -\frac{1}{2\pi v}(2S+1)(g_{j,1 , \parallel})^{2}
\:\:\:\: , 
\label{fer.21}
\end{eqnarray}
\noindent
with  $S=0$ for spinless electrons and $S=1/2$ otherwise \cite{solyom,matgla1,matgla2}.  
Clearly, the renormalization group equations for the $S$-matrix must be 
supplemented with   Eqs. (\ref{eq:fermionic4}) in the spinful case in which, 
differently from what happens in the spinless case, there is a nontrivial 
flow of the bulk interaction parameters with the scale $L / \lambda$. 

In the analysis of our paper, we also have to consider a 
generalized bulk interaction which, for $g_{ j , 1 , \perp} = 0$,
has a nonzero inter-wire component, that is, the generalized bulk
interaction   Hamiltonian reads

\beq
H_{\rm Int} = H_{\rm Intra} + H_{\rm Inter} 
\;\;\;\; ,
\label{fer.a1}
\eneq
\noindent
with 

\begin{eqnarray}
 H_{\rm Intra} &=& \sum_{ j = 1}^K \: \sum_{ \sigma , \sigma'} 
 \: g_{j ; (\sigma , \sigma')}  \int_0^L \: d x \: 
 \rho_{ R , j , \sigma} ( x )  \rho_{ L , j , \sigma'} ( x )
 \nonumber \\
  H_{\rm Inter} &=& \sum_{ j \neq j' = 1}^K \: \sum_{ \sigma , \sigma'} 
 \: g_{(j , j')  ; (\sigma , \sigma')}  \int_0^L \: d x \: 
 \rho_{ R , j , \sigma} ( x )  \rho_{ L , j' , \sigma'} ( x )
 \;\;\;\; , 
 \label{fer.a2}
\end{eqnarray}
\noindent
which reduces to $H_{\rm Int}$ in Eq. (\ref{fer.12}), provided 
one identifies the intra-wire interaction strengths as 

\begin{eqnarray}
 g_{ j ; ( \uparrow , \uparrow)} &=& g_{j ; ( \downarrow , \downarrow)} \equiv 
 g_{ j , 2 , \parallel} - g_{ j , 1 , \parallel}  \nonumber \\
 g_{ j  ; ( \uparrow , \downarrow)} &=& g_{ j  ; ( \downarrow , \uparrow)}
 \equiv  g_{ j , 2 , \perp }
 \:\:\:\: , 
 \label{fer.a3}
\end{eqnarray}
\noindent
and sets $g_{ ( j , j' )  ; ( \sigma , \sigma')} = 0$ for $j \neq j'$. 
On assuming again that the $S$-matrix takes the spin-diagonal form 
in Eq. (\ref{2wj.1}), one finds that the RG-equations for the $S$-matrix
elements are again those provided in Eq. (\ref{fer.14}), but with 
the $F$-matrix elements now given by

\beq
F_{ j , j'} = \{ \delta_{ j , j'} \: \frac{g_{ j ; ( \sigma , \sigma)} }{4 \pi v }
+ [ 1 - \delta_{ j , j'} ] \: \frac{g_{ ( j , j')  ; ( \sigma , \sigma)} }{4 \pi v } \} S_{ j , j'}
\:\:\:\: . 
\label{fer.a4}
\eneq
\noindent
As we discuss in the main text, the modification in Eq. (\ref{fer.a4}) does not
substantially affect the solutions of the RG equations.

 \section{Review of bosonization rules for interacting one-dimensional quantum wires}
 \label{boso_analysis}
 
In this Appendix, we review the basic bosonization formulas for one-dimensional
interacting electronic systems, which provide us with the fundamental formal ground, 
on which we relate when defining the effective fermionic coordinates in the 
strongly interacting regimes. The first step towards a fully bosonic description of 
the junction is to rewrite in bosonic coordinates the  free  Hamiltonian $H_{0 }$ 
in Eq. (\ref{fer.1}). This is done by introducing  $K$ bosonic fields $\Phi_{ c , j}$, described by 
the (charge-sector) Hamiltonian 

\beq
H_{0;B;c } = \frac{1}{4 \pi} \:  \int_0^L  \: d x  \: \sum_{ j = 1}^K 
\left[ \frac{1}{v} ( \partial_t \Phi_{c , j} )^2 + v ( \partial_x \Phi_{c , j} )^2 \right] 
\;\;\;\; , 
\label{an.2}
\eneq
\noindent
and   $K$ bosonic fields $\Phi_{ s , j}$, described by 
the (spin-sector) Hamiltonian 

\beq
H_{0;B;s} = \frac{1}{4 \pi} \:  \int_0^L  \: d x  \: \sum_{ j = 1}^K 
\left[ \frac{1}{v} ( \partial_t \Phi_{ s  , j} )^2 + v ( \partial_x \Phi_{ s , j} )^2 \right] 
\;\;\;\; , 
\label{an.2k}
\eneq
\noindent
together with the corresponding dual fields $\Theta_{c , j} , \Theta_{s , j} $, which are 
related to the $\Phi$-fields by means of 
the cross derivative relations

\begin{eqnarray}
 \partial_t \Phi_{ j , c ( s )}  &=& v \partial_x \Theta_{ j , c ( s )}  
 \nonumber \\
 \partial_t \Theta_{ j , c ( s )}  &=& v \partial_x \Phi_{ j , c ( s )}  
\:\:\:\: . 
\label{an.4}
 \end{eqnarray}
\noindent
Therefore, one rewrites the chiral fermionic fields in terms of 
the bosonic coordinates introduced above as  

\begin{eqnarray}
 \psi_{R , \sigma , j } ( x ) &=& \eta_{R , \sigma , j } 
 e^{ \frac{i}{2} [ \Phi_{ j , c } ( x ) + 
 \Theta_{ j , c } ( x ) + \sigma ( \Phi_{ j , s } ( x ) + 
 \Theta_{ j , s } ( x ) )]} \nonumber \\
  \psi_{L , \sigma , j } ( x ) &=& \eta_{L , \sigma , j } e^{ \frac{i}{2} [  
  \Phi_{ j , c } ( x ) - 
 \Theta_{ j , c } ( x ) + \sigma ( \Phi_{ j , s } ( x ) - 
 \Theta_{ j , s } ( x ) )]} 
 \;\;\;\; , 
 \label{an.3}
\end{eqnarray}
\noindent
with $ \eta_{R , \sigma , j }$ and  $\eta_{L , \sigma , j } $ being
real-fermion Klein factors.  Note that the vertex operators in Eq. (\ref{an.3})
have been set consistently with the fact that, as a consequence of 
Eqs. (\ref{an.4}), one finds that the following linear combinations are chiral fields

\begin{eqnarray}
 \phi_{R , c (s ) , j }( x , t )  &=& \frac{1}{\sqrt{2}} [ \Phi_{ j , c ( s )} ( x , t ) + 
 \Theta_{ j , c ( s )} ( x , t )]  \nonumber \\
  \phi_{L , c (s ) , j }( x , t )  &=& \frac{1}{\sqrt{2}} [ - \Phi_{ j , c ( s )} ( x , t ) + 
 \Theta_{ j , c ( s )} ( x , t )] 
 \;\;\;\; . 
 \label{an.5}
\end{eqnarray}
\noindent
In bosonic coordinates, the chiral fermionic density
operators at fixed spin polarization are realized as 

\begin{eqnarray}
: \psi_{R,\sigma , j}^\dagger ( x ) \psi_{ R , \sigma , j } ( x ) : &=& 
\frac{1}{8 \pi} \{ - \partial_x \Theta_{j , c} ( x ) + \partial_x
\Phi_{ j , c } ( x ) + \sigma [  - \partial_x \Theta_{j , s} ( x ) + \partial_x
\Phi_{ j , s } ( x ) ] \} \nonumber \\
: \psi_{L,\sigma , j}^\dagger ( x ) \psi_{ L , \sigma , j } ( x ) : &=& 
- \frac{1}{8 \pi} \{  \partial_x \Theta_{j , c} ( x ) + \partial_x
\Phi_{ j , c } ( x ) + \sigma [   \partial_x \Theta_{j , s} ( x ) + \partial_x
\Phi_{ j , s } ( x ) ] \}
\:\:\:\: ,
\label{an.9}
\end{eqnarray}
\noindent
with the double columns $ : \: :$ denoting normal-ordering with respect to 
the fermionic groundstate. From Eqs. (\ref{an.9}) one therefore recovers the   
charge- and the spin-density operators given by 

\begin{eqnarray}
\rho_{ c , j } ( x ) = 
\sum_\sigma \{ : \psi_{R , \sigma , j }^\dagger  ( x ) \psi_{R , \sigma , j } ( x ) + 
\psi_{L , \sigma , j }^\dagger  ( x ) \psi_{L , \sigma , j } ( x ) : \} &\to& -
\frac{1}{2 \pi} \: \partial_x \Theta_{ j , c} ( x ) \nonumber \\
\rho_{ s , j } ( x ) = 
\sum_\sigma \sigma \{ : \psi_{R , \sigma , j }^\dagger  ( x ) \psi_{R , \sigma , j } ( x ) + 
\psi_{L , \sigma , j }^\dagger  ( x ) \psi_{L , \sigma , j } ( x ) : \} &\to& -
\frac{1}{2 \pi} \: \partial_x \Theta_{ j , s} ( x )
\:\:\:\: , 
\label{aleph.1}
\end{eqnarray}
\noindent
together with the corresponding current operators  realized as

\begin{eqnarray}
J_{ c , j } ( x ) = 
\sum_\sigma \{ : \psi_{R , \sigma , j }^\dagger  ( x ) \psi_{R , \sigma , j } ( x ) -
\psi_{L , \sigma , j }^\dagger  ( x ) \psi_{L , \sigma , j } ( x ) : \} &\to& 
\frac{1}{2 \pi} \: \partial_x \Phi_{ j , c} ( x ) \nonumber \\
J_{ s , j } ( x ) = 
\sum_\sigma \sigma \{ : \psi_{R , \sigma , j }^\dagger  ( x ) \psi_{R , \sigma , j } ( x ) - 
\psi_{L , \sigma , j }^\dagger  ( x ) \psi_{L , \sigma , j } ( x ) : \} &\to& 
\frac{1}{2 \pi} \: \partial_x \Phi_{ j , s} ( x )
\;\;\;\; . 
\label{an.6}
\end{eqnarray}
\noindent
All the previous transformations lead to contributions to the 
system Hamiltonian that are quadratic in the bosonic fields. 
The interaction Hamiltonian shares this property, once
one has set to zero the  all the terms $\propto g_{j,1 , \perp}$,
in which case, on rewriting  $H_{\rm int}$ in 
Eq. (\ref{eq:fermionic1}) in terms of the bosonic fields, one obtains 
$ H_{\rm int} = H_{\rm int , 1} + H_{\rm int , 2}$, with

\begin{eqnarray}
H_{\rm int , 1} &=&    \sum_{j = 1}^K
\frac{g_{j,1 , \parallel} -  g_{j,2 , \parallel}}{16 \pi^2}
\:  \int_0^L  \: d x \: \{ - ( \partial_x \Theta_{ j , c  } ( x ))^2 - 
 ( \partial_x \Theta_{ j , s } ( x ))^2 +  ( \partial_x \Phi_{ j , c  } ( x ))^2  
+ ( \partial_x \Phi_{ j , s  } ( x ))^2 \} \nonumber \\
 H_{\rm int , 2} &=&    \sum_{j = 1}^K
\frac{g_{j,2 , \perp} }{16 \pi^2} \:  \int_0^L  \: d x \: \{ 
( \partial_x \Theta_{ j , c } ( x ))^2 + ( \partial_x \Phi_{ j , s} ( x ))^2 
- ( \partial_x \Theta_{ j , s } ( x ))^2 -
( \partial_x \Phi_{ j , c } ( x ))^2  \}
\:\:\:\: . 
\label{an.11}
\end{eqnarray}
\noindent
On adding the terms in Eqs. (\ref{an.11}) to 
the noninteracting Hamiltonian for the charge sector (Eq. (\ref{an.2})) 
plus the one for the spin sector (Eq. (\ref{an.2k})), one eventually obtains 
a bosonic Hamiltonian $H$ that is still quadratic, although with 
pertinently renormalized coefficients, given by

\begin{eqnarray}
 H &=& \frac{1}{4 \pi} \: \sum_{ j = 1}^K u_{ j , c} \: 
  \int_0^L \: d x \: \left\{ g_{j,  c  } ( \partial_x \Phi_{ j , c } ( x ))^2 
 + \frac{1}{g_{j , c } } ( \partial_x \Theta_{ j , c } ( x ))^2 \right\}
 \nonumber \\ 
 &+& \frac{1}{4 \pi} \: \sum_{ j = 1}^K u_{ j , s } \: 
 \int_0^L  \: d x \: \left\{ g_{j , s } ( \partial_x \Phi_{ j ,  s  } ( x ))^2 
 + \frac{1}{g_{j,s } } ( \partial_x \Theta_{ j ,  s } ( x ))^2 \right\}
 \:\:\:\: , 
 \label{an.13}
\end{eqnarray}
\noindent
with

\begin{eqnarray}
  u_{ j , c} g_{j,  c } &=& v \left[ 1 + \frac{g_{j , 1 , \parallel} -
  g_{j , 2 , \parallel} - g_{ j , 2 , \perp}}{4 \pi v} \right] \nonumber \\
   \frac{ u_{ j , c } }{  g_{j,  c  } } &=& v \left[ 1 - \frac{g_{j , 1 , \parallel} -
  g_{j , 2 , \parallel} - g_{ j , 2 , \perp}}{4 \pi v} \right] \nonumber \\
    u_{ j , s } g_{j,  s } &=& v \left[ 1 + \frac{g_{j , 1 , \parallel} -
  g_{j , 2 , \parallel} + g_{ j , 2 , \perp}}{4 \pi v} \right] \nonumber \\
   \frac{ u_{ j , s } }{  g_{j,  s  } } &=& v \left[ 1 - \frac{g_{j , 1 , \parallel} -
  g_{j , 2 , \parallel} + g_{ j , 2 , \perp}}{4 \pi v} \right] 
 \:\:\:\: . 
 \label{an.14}
\end{eqnarray}
\noindent
(Note  the use of Eqs. (\ref{an.4}) to express $H$ in terms of both 
the $\Phi$- and the $\Theta$-fields.) When $g_{j,1 , \perp} \neq 0$, 
one can employ the identities

\begin{eqnarray}
 \psi_{R,\sigma , j }^\dagger ( x ) \psi_{ R , \bar{\sigma} , j } ( x ) &\to& 
 e^{- i  \sigma [ \Phi_{ j , s } ( x ) + \Theta_{ j , s } ( x ) ] } \nonumber \\
 \psi_{L , \sigma , j}^\dagger ( x ) \psi_{L , \bar{\sigma} , j } ( x ) &\to& 
  e^{- i  \sigma [ \Phi_{ j , s } ( x ) - \Theta_{ j , s } ( x ) ] }
  \:\:\:\: , 
  \label{an.15}
\end{eqnarray}
\noindent
to express the total additional contribution to $H_{\rm int}$, $H_{\rm int , 3}$, as

\beq
H_{\rm int , 3} \sim  \: \int_0^L \: d x \: \sum_{ j =1}^K \: W_j \: 
\cos [ 2 \Phi_{ j , s } ( x ) ] 
\:\:\:\: ,
\label{an.16}
\eneq
\noindent
with $W_j \propto g_{  j , 1 , \perp}$.
To lowest order, the renormalization group equation for $W_j$ is

\beq
\frac{d W_j}{d \ell} = \left[ 2 -  \frac{2}{ g_{j , s } } \right] W_j
\:\:\:\: ,
\label{an.17}
\eneq
\noindent
which   tells us that  
the nonlinear interaction term $\propto g_{  j , 1 , \perp}$
is  irrelevant as long as $g_{j , s }  < 1$ and, accordingly, 
even if  $g_{  j , 1 , \perp}$ has not been fine-tuned to 0, 
it can be dropped out of the effective low-energy, long-wavelength 
theory. In general, when using the bosonization, we assume that either this is the case, or 
that all the $g_{  j , 1 , \perp}$ have been fine-tuned to 0. As for what 
concerns the boundary Hamiltonian $H_B$ describing the junction between the 
quantum wires, throughout all the paper we have assumed that total spin
was conserved at each scattering event at the junction. This basically 
implies that $H_B$ can be fully as in Eq. (\ref{th.1}) of the main text. 
While for a weak bulk interaction within the wires 
the most relevant contribution to 
$H_B$ is typically realized as a linear combinations of (some of) the 
operators in Eq. (\ref{appe.bou1}), at strong enough attraction either in 
the charge-, or in the spin-channel (or both), as we discuss in the paper, 
the most relevant boundary interaction term can be realized as 
a product of two $B$-operators. In any case, all the boundary operators can 
be rewritten in bosonic coordinates by pertinently employing 
the bosonization rules in Eqs. (\ref{an.9}). As the junction naturally
defines a (common) boundary for all the quantum wires, Eqs. (\ref{an.9}) 
must typically be supplemented with  the appropriate
boundary conditions for the $\Phi_j$- and for the $\Theta_j$-fields. 
This ''delayed evaluation of the boundary conditions'' (DEBC)-procedure 
can be typically implemented in the correspondence of a conformal
fixed point \cite{oca1,oca2}. It has the net effect of making 
different $B$-operators in Eq. (\ref{appe.bou1})
to ''collapse'' onto each other, thus substantially reducing the 
number of independent boundary operators that can potentially enter 
$H_B$ and, thus, strongly constraining the number of relevant scattering 
processes that can take place in the proximity of a given fixed point.
For instance,   the fields $\Phi_{ j , c (s ) } ( x )$ corresponding to a wire that is disconnected
from the junction must obey Neumann boundary conditions such as 
$\partial_x \Phi_{ j , c (s) } ( x =0 ) =0$ and, correspondingly, 
the fields $\Theta_{ j , c (s) } ( x )$ must
be pinned at $x  = 0$ (Dirichlet boundary conditions). Such boundary conditions 
only allow for nontrivial boundary operators in the form 
$B_{ ( j , j ' ) , \sigma , (R , L ) } ( 0 ) \sim \exp \left[ - \frac{i}{2} ( 
\Phi_{ c , j } ( 0 )  - \Phi_{ c , j'} ( 0 ) ) - \frac{i \sigma }{2} ( 
\Phi_{ s , j } ( 0 )  - \Phi_{ s , j'} ( 0 ) )  \right]$ (plus the corresponding 
Hermitean conjugates). Various alternative possible situations are discussed in the 
paper.

\section{Linear response theory approach to the conductance tensor of 
a junction of quantum wires}
\label{ltr}

In this section, we review the linear response theory approach to the 
derivation of the conductance tensor for a junction of quantum wires. 
After developing the main formula for computing the conductance tensor within 
linear response theory, we will apply it to a number specific cases we analyze in 
the main text of our paper. 

Let ${\cal W}_j$ be the $j$-th wire connected to the junction ($j = 1 , \ldots , K$).  
In order to implement linear response theory, we imagine to connect 
 ${\cal W}_j$ to a reservoir ${\cal R}_j$, which can either be characterized 
by the same parameters as the wire to which it is connected, or not 
(a typical situation corresponds to ${\cal W}_j$ being an interacting 
one-dimensional quantum wire, described as a single Luttinger liquid, connected 
to a Fermi liquid reservoir \cite{maslov_stone}). For the sake of simplicity, 
we assume that the parameters characterizing both the wires and the reservoirs are all 
independent of $j$.  
In order to induce a current flow across the junction, we assume that 
each reservoir is characterized by an equilibrium distribution for particles with 
spin $\sigma$, with  
chemical potential $\mu_{j , \sigma}  = e V_{ j , \sigma} $. This induces electric fields $\{ {\cal E}_{j , \sigma} ( t ) \}$
distributed  in the various branches of the junction, which we account for by introducing
a set of uniform vector potentials ${\cal A}_{j , \sigma}  ( t )$, one for each wire, 
such that ${\cal E}_{j , \sigma}  ( t ) = - \partial_t  {\cal A}_{ j , \sigma}  ( t )$.  Letting $J_{j , \sigma} ( x )$ 
be the current operator for particles with spin $\sigma$ in wire-$j$ and letting  each
wire to be of length $L$, we may define  the ''source'' Hamiltonian $H_{\rm Source} ( t )$, describing 
the coupling to the applied electric fields, given by

\beq
H_{\rm Source} ( t ) = \sum_{ j = 1}^K \: \sum_\sigma \: \int_\delta^L \: dx  \: 
{\cal A}_{ j , \sigma}  ( t ) J_{j , \sigma}  ( x )
\:\:\:\: . 
\label{applr.1}
\eneq
\noindent
Let us, now, denote with $J_{ (j , \sigma)  ; I } ( x , t ) $, 
$H_{{\rm Source} ; I } ( t )$ the operators taken in the interaction 
representation with respect to the Hamiltonian without the source term.   
Within linear response theory, the current of particles with spin $\sigma$
evaluated at point $x$ of wire-$j$ 
is therefore given by

\beq
I_{j , \sigma}  ( x , t ) = i \sum_{ j' = 1}^K \:\sum_{ \sigma'} \:  \int_{ - \infty}^\infty \: d t' \: 
\: \int_\delta^L \: dx'  \: {\cal D}_{ (j , \sigma )  ;  ( j' , \sigma') } ( x , t ; x' , t' ) 
{\cal A}_{j' , \sigma'} ( t' ) 
\:\:\:\: , 
\label{applr.2}
\eneq
\noindent
with 

\beq
 {\cal D}_{ (j , \sigma)  ;  ( j' , \sigma') } ( x , t ; x' , t' ) = 
\theta ( t - t' ) \langle [ J_{ (j , \sigma)  ; I } ( x , t ) , 
 J_{ ( j' , \sigma' )  ; I } ( x' , t' )  ] \rangle 
 \:\:\:\: . 
 \label{applr.3}
 \eneq
 \noindent
Equation (\ref{applr.2}) does generically apply to any situation, whether
the wires are interacting, or not, and whether one uses a fermionic, or 
a bosonic representation for the Hamiltonian of the junction. An important 
remark, however, is that, in any case, the current must be consistently 
probed outside of the region across which the electric fields are 
applied, that is, in Eq. (\ref{applr.2}) one has always to assume 
that $x > L$. Another important formula is the Fourier-space 
counterpart of Eq. (\ref{applr.2}), that is

\beq
I_{j , \sigma}  ( x , \omega ) = - \frac{1}{\omega} \sum_{ j' = 1}^K \: \sum_{\sigma'} \:  
\: \int_\delta^L \: dx'  \: {\cal D}_{ (j , \sigma ) ; ( j' , \sigma') } ( x , x' ; \omega  ) 
{\cal E}_{j' , \sigma' } (  \omega )
\;\;\;\; , 
\label{applr.4}
\eneq
\noindent
with 

\begin{eqnarray}
 I_{j , \sigma}  ( x , \omega ) &=& \int \: d t \: e^{ i \omega t}  I_{j , \sigma}  ( x ,t )
 \nonumber \\
 {\cal E}_{j' , \sigma'} ( \omega ) &=&   \int \: d t \: e^{ i \omega t}  {\cal E}_{j' , \sigma'} ( t )
 \nonumber \\
  {\cal D}_{ ( j  ,  \sigma) ; ( j' , \sigma' ) } ( x , x' ; \omega  )  &=&   \int \: d t \: e^{ i \omega  t  } 
   {\cal D}_{ ( j , \sigma) ; ( j' , \sigma' ) } ( x , t ; x' , 0 )
   \:\:\:\: . 
   \label{applr.5}
\end{eqnarray}
\noindent
In the following part of this Appendix, we  consider some specific applications to various cases 
we analyze in the main text. An important observation to make at this point is that, on connecting 
the wires to the reservoirs, we basically assume a continuity condition for the current operator
at the interface. From the microscopical point of view, this appears to be the ''macroscopic'' 
counterpart of the ''smooth'' crossover in the interaction strength in real space discussed 
in the microscopic lattice model considered in Ref. [\onlinecite{meden1}]. 
It would be also interesting to work out a macroscopic field-theoretical 
Hamiltonian describing a ''sharp'' interface in the microscopic model, but this 
goes beyond of the scope of this work. As specified above, in the following we 
assume current continuity at the interface, that is, a smooth crossover in the 
bulk interaction strength.

\subsection{Junction of noninteracting quantum wires}
\label{junoninter}

\begin{figure}
\includegraphics*[width=.5\linewidth]{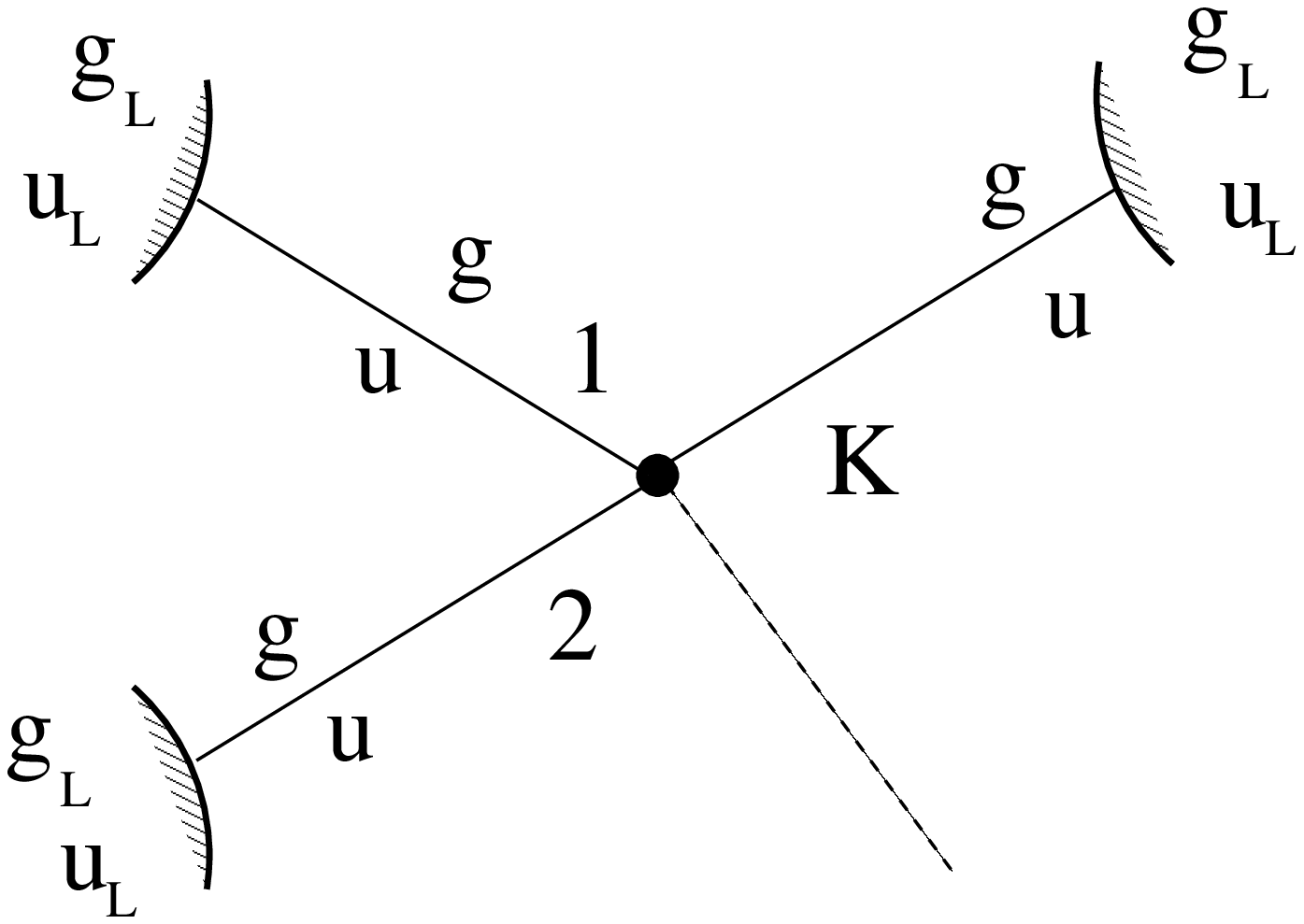}
\caption{Sketch of a connected junction of $K$ wires with 
parameters $g , u$ connected to reservoirs with parameters
$g_L , u_L$.} \label{junction_1}
\end{figure}

Due to the absence of multi-particle processes, 
the conductance tensor for a  junction of noninteracting quantum wires 
can be fully expressed in terms  of  only single-fermion $S$-matrix elements. 
The bulk Hamiltonian for a noninteracting junction of $K$ quantum wires is 
given in Eq. (\ref{fer.1}). Within each wire the chiral fields can be 
expressed in terms of their Fourier modes as 

\begin{eqnarray}
 \psi_{R , j , \sigma  }( x ) &=& \frac{1}{\sqrt{L}} \; \sum_k \; e^{ i k x } a_{R , j , \sigma } ( k ) \nonumber \\
 \psi_{L , j  . \sigma }( x ) &=& \frac{1}{\sqrt{L}} \; \sum_k \; e^{ - i k x } a_{L , j , \sigma  } ( k )
 \;\;\;\; , 
 \label{lrt.2}
\end{eqnarray}
\noindent
with the right-handed and the left-handed chiral modes related to 
each other by the $S$-matrix elements as

\beq
a_{R , j  , \sigma } ( k ) = \sum_{ j' = 1}^K \: \sum_{ \sigma'} 
\; S_{ (j , \sigma ) ; ( j' , \sigma' ) } ( k ) a_{L , j' , \sigma' } ( k )
\;\;\;\; . 
\label{lrt.2x}
\eneq
\noindent
The current operator in wire-$j$ for particles with spin polarization $\sigma$ is 
given by

\beq
J_{ j , \sigma } ( x ) = e v \{ : \psi_{ R , j , \sigma}^\dagger ( x ) \psi_{ R , j , \sigma } ( x ) : 
- : \psi_{ L , j , \sigma}^\dagger ( x ) \psi_{ L , j , \sigma } ( x ) :  \}
\:\:\:\: , 
\label{lrt.a2}
\eneq
\noindent
which, taking into account that the $S$-matrix is diagonal in the spin indices,  implies

\beq
{\cal D}_{ ( j , \sigma ) ; (  j' , \sigma')  } ( x , x' ;  \omega ) = e^2 v^2 \delta_{ \sigma , \sigma'} 
\{  {\cal D}_{(R,R); j , j' } ( x ,  x' ;  \omega ) + 
{\cal D}_{(L,L); j , j' } ( x ,  x' ;  \omega ) - 
{\cal D}_{(R,L); j , j' } ( x ,  x' ;  \omega ) - 
{\cal D}_{(L,R); j , j' } ( x ,  x' ;  \omega ) \}
\:\:\:\: . 
\label{lrt.9}
\eneq
\noindent
Assuming a thermal distribution for 
fermions in the reservoirs, one therefore obtains 

\begin{eqnarray}
 {\cal D}_{(R,R); j , j' } ( x ,  x' ;  \omega ) &=& \frac{i \delta_{ j , j'} }{4 \pi^2 v^2}
 \: \int \: d E \: d E' \: f ( E ) f ( E' ) \left[ 
 \frac{e^{ - i \left( \frac{E + E'}{v} \right) ( x -  x' )} }{E' + E + \omega
 + i \eta} + \frac{e^{  i \left( \frac{E + E'}{v} \right) ( x -  x' )} }{E' + E -  \omega
 - i \eta}  \right] \nonumber \\
  {\cal D}_{(L,L); j , j' } ( x ,  x' ;  \omega ) &=& \frac{i \delta_{ j , j'} }{4 \pi^2 v^2}
 \: \int \: d E \: d E' \: f ( E ) f ( E' ) \left[ 
 \frac{e^{  i \left( \frac{E + E'}{v} \right) ( x -  x' )} }{E' + E + \omega
 + i \eta} + \frac{e^{ - i \left( \frac{E + E'}{v} \right) ( x -  x' )} }{E' + E -  \omega
 - i \eta}  \right]  \nonumber \\
  {\cal D}_{(R,L); j , j' } ( x ,  x' ;  \omega ) &=& \frac{i | S_{ j , j'} |^2 }{4 \pi^2 v^2}
 \: \int \: d E \: d E' \: f ( E ) f ( E' ) \left[ 
 \frac{e^{ - i \left( \frac{E + E'}{v} \right) ( x +  x' ) }}{E' + E + \omega
 + i \eta} + \frac{e^{  i \left( \frac{E + E'}{v} \right) ( x +  x' )} }{E' + E -  \omega
 - i \eta}  \right]
 \nonumber \\
   {\cal D}_{(L,R); j , j' } ( x ,  x' ;  \omega ) &=& \frac{i | S_{ j' , j} |^2 }{4 \pi^2 v^2 }
 \: \int \: d E \: d E' \: f ( E ) f ( E' ) \left[ 
 \frac{e^{  i \left( \frac{E + E'}{v} \right) ( x +  x')} }{E' + E + \omega
 + i \eta} + \frac{e^{  - i \left( \frac{E + E'}{v} \right) ( x +  x' )} }{E' + E -  \omega
 - i \eta}  \right]
 \;\;\;\; , 
 \label{lrt.10}
\end{eqnarray}
\noindent
with $ S_{ j , j' }$ denoting the $S$-matrix elements computed at the Fermi level, 
$f ( E)$ being the Fermi distribution function,  and $\eta = 0^+$. Since, as stated 
above, we assume that the current is measured outside of the region over which the electric
fields are applied, in the following we  assume $x , x' > 0 $, $0 < x' < L$ and $x > L$. 
As a result, integrating in Eqs. (\ref{lrt.10}) over $d E'$ and closing the integration path over the 
appropriate half-plane   according to what is the  phase factor in the 
argument of the integrals, one   obtains 

\begin{eqnarray}
  {\cal D}_{(R,R); j , j' } ( x ,  x' ;  \omega ) &=&  \frac{ \delta_{j , j'} }{2 \pi v^2}
  \: \int \: d E \: e^{ i \frac{\omega}{v} ( x -  x' ) } f ( E ) 
  \{ f ( - E - \omega ) - f ( - E + \omega ) \} \nonumber \\
   {\cal D}_{(L,L); j , j' } ( x ,  x' ;  \omega ) &=& 0 \nonumber \\
      {\cal D}_{(R,L); j , j' } ( x ,  x' ;  \omega ) &=& 
      \frac{ | S_{j , j'} |^2 }{2 \pi v^2}
  \: \int \: d E \: e^{ i \frac{\omega}{v} ( x +  x' ) } f ( E ) 
  \{ f ( - E - \omega ) - f ( - E + \omega ) \} \nonumber \\
   {\cal D}_{(L,R); j , j' } ( x ,  x' ;  \omega ) &=& 0 
   \:\:\:\: . 
   \label{lrt.11}
\end{eqnarray}
\noindent
Going to the zero-$T$ limit and focusing onto 
the dc current ($\omega \to 0$), one recovers that the current in wire-$j$ for particles
with spin polarization $\sigma$, $I_{ j , \sigma }$, is given by
limit

\beq
I_{ j , \sigma }  = \frac{e^2}{2 \pi} \: \sum_{ j' = 1}^K \{- | S_{ j , j'} |^2 + \delta_{ j , j' } \} 
V_{j' , \sigma } 
\;\;\;\; .
\label{lrt.12}
\eneq
\noindent
The dc-conductance matrix elements $G_{ ( j , \sigma ) ; ( j' , \sigma' ) }$ are therefore 
given by

\beq
G_{ ( j , \sigma ) ; ( j' , \sigma' ) } = \frac{ \partial I_{ j , \sigma }}{\partial 
V_{j' , \sigma } } = \delta_{ \sigma , \sigma'} \frac{e^2}{2 \pi} 
\: \{- | S_{ j , j'} |^2 + \delta_{ j , j' } \}  
\:\:\:\: , 
\label{lrt.12b}
\eneq
\noindent
The charge- and the spin-conductance tensors $G_c , G_s$, are respectively defined as 

\beq
G_{ c ; ( j , j' )} = \sum_{ \sigma , \sigma'} G_{ ( j , \sigma ) ; ( j' , \sigma' ) } 
\;\;\; , \;\; 
G_{ s ; ( j , j' )} = \sum_{ \sigma , \sigma'} \sigma \sigma' G_{ ( j , \sigma ) ; ( j' , \sigma' ) } 
\;\;\;\: , 
\label{lrt.12c}
\eneq
\noindent
which, from Eq. (\ref{lrt.12b}), implies 

\beq
G_{ c ; ( j , j' )} = G_{ s ; ( j , j' )} =  \frac{e^2}{  \pi} 
\: \{- | S_{ j , j'} |^2 + \delta_{ j , j' } \} 
\:\:\:\: . 
\label{lrt.12d}
\eneq
\noindent
When a weak bulk interaction is added to the junction Hamiltonian, using the FRG-approach 
as  we do in  Sec. \ref{three_wire}, we   still use Eq. (\ref{lrt.12d}) for the 
charge- and the spin-conductance tensors by just replacing the ''bare'' $S$-matrix elements
with the running ones, $S_{ j , j' } ( D )$. For instance, this led us to 
write Eq. (\ref{cflow}) of the main text.  The result one obtains does actually 
correspond to the conductance tensor of the ''connected'' junction, 
that is, of the junction of weakly interacting wires connected to 
bulk Fermi liquid reservoirs. As pointed 
out in \cite{maslov_stone,oca1,oca2}, the tensor $G_{c (s) ; (j , j') }$ defined in Eq. (\ref{lrt.12b}) are  related 
to their analogs for the junction disconnected from the leads,   $\bar{G}_{ c ( s) ; ( j,j')}$,  
by means of pertinent generalizations of Eq. (2.7) of  Refs. [\onlinecite{oca1,oca2}], that is, 

\beq
[ G^{-1} ]_{ c(s) ; ( j , j' ) } = [ \bar{G}^{-1} ]_{ c (s) ; ( j , j' ) } + 
 G_{{\rm in} , c(s) }^{-1} \delta_{j,j'} 
\:\:\:\: , 
\label{ltr.13}
\eneq
\noindent
with the interface charge (spin) conductance $G_{{\rm in} , c(s) } = \frac{2 g_{c(s) }}{g_{c(s)} - 1}
\frac{e^2}{ \pi}$. From Eq. (\ref{ltr.13}) one sees that the matrix elements 
$G_{ c(s) ; ( j , j' )}$ 
explicitly depend on the bulk interaction parameters via the Luttinger parameters $g_{ c (s)}$
(see Fig. \ref{junction_1} for a sketch of the junction connected to reservoirs with 
generic values of the Luttinger parameter $g_L$). 
Equation (\ref{ltr.13}) suggests how to recover the conductance tensor 
for the junction disconnected from the leads within the FRG-approach: it is enough 
to just compute the $G_{ c(s) ; ( j , j')}$s
and, therefore, to invert Eq. (\ref{ltr.13}) to derive the $\bar{G}_{ c(s) ; ( j , j')}$s.

We now discuss how Eq. (\ref{ltr.13}) is modified in the 
dual models we used to apply the FRG-approach to junctions at a 
strong bulk interaction in the wire. 
 
\noindent
 
\subsection{Junction of strongly interacting quantum wires at $g_c = g_s = 3$}
\label{appe33}

As Eq. (\ref{applr.4}) applies independently of the specific form of the 
current operators, in order to generalize  Eq. (\ref{ltr.13}) to the strongly
interacting case we discuss in Sec. \ref{gg33}, one has just to 
compute the ${\cal D}_{ (j , \sigma)  ;  ( j' , \sigma') } ( x , t ; x' , t' ) $-terms
in Eq. (\ref{applr.3}) using the current operators in the dual theory given in Eq. (\ref{strint.4h}).
As a result, one now obtains 
  ${\cal D}_{ ( j  , \sigma ) ; ( j' , \sigma') } ( x, x' ; \omega ) 
 = \delta_{ \sigma , \sigma'} {\cal D}_{ j , j'} ( x , x' ; \omega )$, with 
 
 \begin{eqnarray}
  {\cal D}_{ j , j'} ( x , x' ; \omega ) &=& \frac{e^2}{2 \pi} \: \int \: d E \: 
  f ( E ) [ f ( - E - \omega ) - f ( - E + \omega ) ] \nonumber \\
  &\times & \Biggl\{ 2 e^{ i \frac{\omega}{u} ( x - x' ) } \delta_{ j , j'} 
  + e^{ i \frac{\omega}{u} ( x + x' ) } [ | \tilde{S}_{ j - 1 , j' - 1} |^2 + 
  | \tilde{S}_{ j + 1 , j' + 1} |^2 ] \nonumber \\
  &-&    e^{ i \frac{\omega}{u} ( x - x' ) } [ \delta_{ j - 1 , j' + 1} + 
  \delta_{ j + 1 , j' - 1} ] 
  - e^{ i \frac{\omega}{u} ( x + x' ) } [ | \tilde{S}_{ j - 1 , j' + 1} |^2 + 
  | \tilde{S}_{ j + 1 , j' - 1} |^2 ] \Biggr\}
  \:\:\:\: .
  \label{33a.2}
 \end{eqnarray}
\noindent
 Following the same steps as 
in Appendix \ref{junoninter}, from Eq. (\ref{33a.2}) one therefore readily derives the dc conductance 
tensor, which is given by

\beq
G_{ ( j , \sigma ) ; ( j' , \sigma' ) } = \delta_{ \sigma , \sigma'} 
G_{ j ,j'}
\:\:\:\: , 
\label{33a.3}
\eneq
\noindent
with 

\begin{eqnarray}
G_{ j , j'} &=& \frac{e^2}{2 \pi} \Biggl\{ 2 \delta_{ j ,j'} + | \tilde{S}_{ j  + 1 , j'+1} |^2 + 
| \tilde{S}_{ j - 1 , j' - 1} |^2  \nonumber \\
&-& [ \delta_{ j +1 , j' - 1} + \delta_{ j - 1 , j' + 1} + 
| \tilde{S}_{ j -1 , j'+1} |^2 + 
| \tilde{S}_{ j + 1 , j' - 1} |^2   ]
\Biggr\}
\:\:\:\: . 
\label{33a.4}
\end{eqnarray}
\noindent
From  Eq. (\ref{33a.4}) one therefore computes  the conductance
tensor at $g_c = g_s = 3$. When either 
$g_c$, or $g_s$ (or both) are different from 3, using Eq. (\ref{33a.4}) 
one recovers the  conductance tensor in the case in which $g_c$ and $g_s$ both 
asymptotically tend to 3 in the leads. The charge- and the spin-conductance 
tensors for the disconnected junction are obtained using Eq. (\ref{ltr.13})
with  $G_{{\rm in} , c(s) } = \frac{2 g_{c(s) }}{g_{c(s)} - 1}
\frac{e^2}{ \pi}$.

\bibliography{biblio_new.bib}
\end{document}